\documentclass[twocolumn]{emulateapj}

\usepackage{subfigure}

\usepackage{bm}

\usepackage{amsmath, amssymb}
\usepackage{times}
\usepackage{multirow}

\usepackage[usenames]{xcolor}
\definecolor{mpl_red}{HTML}{D62728}

\usepackage[backref, breaklinks, plainpages=false, colorlinks=true, anchorcolor=blue!50!black, citecolor=blue!50!black, linkcolor=blue!50!black, urlcolor=mpl_red, bookmarks=false]{hyperref}
\usepackage{natbib}
\usepackage[strict]{changepage}

\usepackage{stackengine}
\newcommand\xrowht[2][0]{\addstackgap[.5\dimexpr#2\relax]{\vphantom{#1}}}

\newcommand\extrafootertext[1]{%
	\bgroup
	\renewcommand\thefootnote{\fnsymbol{footnote}}%
	\renewcommand\thempfootnote{\fnsymbol{mpfootnote}}%
	\footnotetext[0]{#1}%
	\egroup
}

\usepackage{placeins}

\usepackage{lipsum,atbegshi,etoolbox}
\makeatletter
\newcommand{\discardpages}[1]{
	\xdef\discard@pages{#1}
	\AtBeginShipout{
		\renewcommand*{\do}[1]{
			\ifnum\value{page}=##1\relax%
			\AtBeginShipoutDiscard
			\gdef\do####1{}
			\fi%
		}%
		\expandafter\docsvlist\expandafter{\discard@pages}
	}%
}
\newif\ifkeeppage
\newcommand{\keeppages}[1]{
	\xdef\keep@pages{#1}
	\AtBeginShipout{
		\keeppagefalse%
		\renewcommand*{\do}[1]{
			\ifnum\value{page}=##1\relax%
			\keeppagetrue
			\gdef\do####1{}
			\fi%
		}%
		\expandafter\docsvlist\expandafter{\keep@pages}
		\ifkeeppage\else\AtBeginShipoutDiscard\fi
	}%
}
\makeatother

\begin{document}

\renewcommand*{\backref}[1]{[#1]}

\newcommand{\Pearlman}{\href{https:/orcid.org/0000-0002-8912-0732}{\textcolor{blue!50!black}{Aaron~B.~Pearlman}}}
\newcommand{\Majid}{\href{https:/orcid.org/0000-0002-4694-4221}{\textcolor{blue!50!black}{Walid~A.~Majid}}}
\newcommand{\Prince}{\href{https:/orcid.org/0000-0002-8850-3627}{\textcolor{blue!50!black}{Thomas~A.~Prince}}}
\newcommand{\Nimmo}{\href{https://orcid.org/0000-0003-0510-0740}{\textcolor{blue!50!black}{Kenzie~Nimmo}}}
\newcommand{\Hessels}{\href{https://orcid.org/0000-0003-2317-1446}{\textcolor{blue!50!black}{Jason~W.~T.~Hessels}}}
\newcommand{\Naudet}{\href{https://orcid.org/0000-0001-6898-0533}{\textcolor{blue!50!black}{Charles~J.~Naudet}}}
\newcommand{\Kocz}{\href{https:/orcid.org/0000-0003-0249-7586}{\textcolor{blue!50!black}{Jonathon~Kocz}}}

\newcommand{\CaltechPhysics}{Division of Physics, Mathematics, and Astronomy, California Institute of Technology, Pasadena, CA 91125, USA; \textcolor{blue}{aaron.b.pearlman@caltech.edu}}
\newcommand{\JPL}{Jet Propulsion Laboratory, California Institute of Technology, Pasadena, CA 91109, USA}
\newcommand{\UA}{Anton Pannekoek Institute for Astronomy, University of Amsterdam, Science Park 904, 1098 XH Amsterdam, The Netherlands}
\newcommand{\ASTRON}{ASTRON, Netherlands Institute for Radio Astronomy, Oude Hoogeveensedijk 4, 7991 PD Dwingeloo, The Netherlands}
\newcommand{\Berkeley}{Department of Astronomy, University of California, Berkeley, CA 94720, USA}
\newcommand{\NDSEG}{$^{\text{6}}$~NDSEG Research Fellow.}
\newcommand{\NSF}{$^{\text{7}}$~NSF Graduate Research Fellow.}


\journalinfo{{\sc Accepted for publication in The Astrophysical Journal Letters, 2020 November 12}}
\submitted{Accepted for publication in The Astrophysical Journal Letters on 2020 November 12}

\shorttitle{MULTIWAVELENGTH RADIO OBSERVATIONS OF FRB~121102 AND FRB~180916.J0158+65}
\shortauthors{PEARLMAN ET AL.}

\title{Multiwavelength Radio Observations of Two Repeating Fast Radio Burst Sources:\\FRB~121102 and FRB~180916.J0158+65}

\author{\Pearlman\altaffilmark{1,6,7}, \Majid\altaffilmark{2,1}, \Prince\altaffilmark{1,2}, \Nimmo\altaffilmark{3,4}, \Hessels\altaffilmark{3,4}, \Naudet\altaffilmark{2}, and \Kocz\altaffilmark{1,5}}

\address{
$^{\text{1}}$~\CaltechPhysics \\
$^{\text{2}}$~\JPL \\
$^{\text{3}}$~\UA \\
$^{\text{4}}$~\ASTRON \\
$^{\text{5}}$~\Berkeley}

\thanks{\NDSEG}
\thanks{\NSF}


\setcounter{footnote}{5}

\begin{abstract}
\label{Section:Abstract}

The spectra of fast radio bursts~(FRBs) encode valuable information about the source's local environment, underlying emission mechanism(s), and the intervening media along the line of sight. We present results from a long-term multiwavelength radio monitoring campaign of two repeating FRB sources, FRB~121102 and FRB~180916.J0158+65, with the NASA Deep Space Network~(DSN) 70-m radio telescopes (DSS-63 and DSS-14). The observations of FRB~121102 were performed simultaneously at 2.3 and 8.4\,GHz, and spanned a total of 27.3\,hr between 2019~September~19 and 2020~February~11. We detected 2~radio bursts in the 2.3\,GHz frequency band from FRB~121102, but no evidence of radio emission was found at 8.4\,GHz during any of our observations. We observed FRB~180916.J0158+65 simultaneously at 2.3 and 8.4\,GHz, and also separately in the 1.5\,GHz frequency band, for a total of 101.8\,hr between 2019~September~19 and 2020~May~14. Our observations of FRB~180916.J0158+65 spanned multiple activity cycles during which the source was known to be active and covered a wide range of activity phases. Several of our observations occurred during times when bursts were detected from the source between 400--800\,MHz with the Canadian Hydrogen Intensity Mapping Experiment~(CHIME) radio telescope. However, no radio bursts were detected from FRB~180916.J0158+65 at any of the frequencies used during our observations with the DSN~radio telescopes. We find that FRB~180916.J0158+65's apparent activity is strongly frequency-dependent due to the narrowband nature of its radio bursts, which have less spectral occupancy at high radio frequencies~($\gtrsim$\,2\,GHz). We also find that fewer or fainter bursts are emitted from the source at high radio frequencies. We discuss the implications of these results on possible progenitor models of repeating FRBs.

~\vspace{-0.2cm}

\noindent
\textit{Unified Astronomy Thesaurus concepts:}~\href{http://astrothesaurus.org/uat/1339}{\textcolor{blue}{Radio bursts~(1339)}}; \href{http://astrothesaurus.org/uat/2008}{\textcolor{blue}{Radio transient sources~(2008)}}

\end{abstract}


\section{Introduction}
\label{Section:Introduction}

\setcounter{footnote}{7}

Fast radio bursts~(FRBs) are transient pulses of radio emission (see~\citealt{Cordes+2019} and~\citealt{Petroff+2019} for recent reviews) that have observed temporal widths ranging from microseconds to milliseconds (e.g., see~\citealt{Cho+2020, Nimmo+2020}) and fluences between $\sim$0.01--1,000\,Jy\,ms (e.g., see~\citealt{Shannon+2018}). Their progenitors are mostly believed to be located at extragalactic distances since the observed dispersion measures~(DMs) of their radio bursts exceed the expected contribution from the column density of Galactic free electrons along the line of sight. The extragalactic nature of~FRBs was definitively confirmed through the sub-arcsecond localization of radio bursts from the first repeating~FRB, FRB~121102~(also referred to as FRB~20121102A;~\citealt{Spitler+2014, Spitler+2016}), to a low-metallicity star-forming dwarf galaxy at a redshift of $z$\,$=$\,0.19~\citep{Chatterjee+2017, Marcote+2017, Tendulkar+2017}. Over 100 FRBs have been published to date (see~\citealt{Petroff+2016} for a catalog\footnote{See \href{http://frbcat.org}{http://frbcat.org.}}), which includes 23~repeating FRB~sources~\citep{Spitler+2016, CHIME+2019a, CHIME+2019b, Kumar+2019, Bochenek+2020, CHIME+2020a, Kumar+2020, Luo+2020}. The volumetric occurrence rates of~FRBs that have not been observed to repeat thus far suggest that a large fraction of these sources should emit multiple bursts over their lifetimes~\citep{Ravi2019}.

A total of 13~extragalactic FRB sources have now been localized to host galaxies at redshifts of 0.034--0.66~\citep{Chatterjee+2017, Marcote+2017, Bannister+2019, Prochaska+2019, Ravi+2019a, Bhandari+2020, Heintz+2020, Law+2020, Macquart+2020, Marcote+2020}, which has demonstrated that FRBs can produce radio bursts with a wide range of luminosities from diverse host galaxies and local environments. Radio bursts from repeating FRBs often exhibit hallmark features that typically distinguish their emission from that of apparently non-repeating sources. Repeating FRBs tend to have larger burst widths, on average, compared to non-repeating~FRBs~\citep{Scholz+2016, CHIME+2019b, Fonseca+2020}, as well as bursts with linear polarization fractions approaching 100\% and a flat polarization position angle~(PA) across their burst envelopes~(e.g., see~\citealt{Michilli+2018a, CHIME+2019b, Chawla+2020, Day+2020, Fonseca+2020}). In some cases, they can also emit bursts with subpulses that drift downwards in frequency with time, which has been dubbed the ``sad trombone'' effect~\citep{CHIME+2019a, CHIME+2019b, Hessels+2019, Josephy+2019}. While these properties suggest that the emission mechanisms and/or local environments of repeating and non-repeating FRB sources may be different, it is not yet clear whether they have different physical origins.

Numerous theoretical models have been proposed to explain the emission behavior of FRBs (see~\citealt{Platts+2019} for an overview\footnote{See \href{https://frbtheorycat.org}{https://frbtheorycat.org.}}). Many of these models invoke coherent radiation mechanisms from compact objects, such as young neutron stars or magnetars (e.g., see~\citealt{Beloborodov2017, Metzger+2017, Lu+2018, Ioka+2020}). Recently, progenitor models involving magnetars have garnered considerable attention thanks to the discovery of an unusually bright millisecond-duration radio burst from the Galactic magnetar, SGR~1935+2154~\citep{Bochenek+2020, CHIME+2020a}. This discovery has demonstrated that extragalactic magnetars are responsible for at least some fraction of the cosmological FRB~population and has also helped to bridge the large radio energy gap that previously existed between Galactic magnetars and FRBs. In fact, SGR~1935+2154 has emitted radio bursts spanning 7~orders of magnitude in luminosity~\citep{Bochenek+2020, CHIME+2020a, Kirsten+2020, Zhang+2020}, ranging from ``normal'' radio bursts from magnetars (e.g., see~\citealt{Pearlman+2018, Pearlman+2020}) to within $\sim$1 order of magnitude of the faintest known FRBs. However, the burst repetition rates and energetics of most active, extragalactic repeating FRB~sources indicate that their progenitors are somehow different from the population of known Galactic magnetars. The volume density of active repeating FRBs is also much smaller than the volume density of Galactic magnetars, even if one assumes that active repeating sources are produced by younger versions of Galactic magnetars that are presumed to have larger magnetic fields and higher activity levels. This implies that the progentiors of repeating~FRBs must be volumetrically rare~\citep{Lu+2020a, Margalit+2020}.

Over the past few years, daily radio observations of the northern hemisphere sky in the 400--800\,MHz frequency band with the Canadian Hydrogen Intensity Mapping Experiment~(CHIME) transit radio telescope have led to discovery of many new repeating~FRB sources~\citep{CHIME+2019a, CHIME+2019b, Fonseca+2020}, enabled by the instrument's large instantaneous field of view~(FoV), wide bandwidth, and high sensitivity~\citep{CHIME+2018}. In particular, the discovery of FRB~180916.J0158+65 (also referred to as FRB~20180916B;~\citealt{CHIME+2019b}), its subsequent localization to a nearby massive spiral galaxy~\citep{Marcote+2020}, and the detection of a 16.35\,d periodicity (or possibly a higher frequency alias of this period) in the burst arrival times~\citep{CHIME+2020b} has facilitated detailed studies of the source via follow-up observations across multiple wavelengths~(e.g., see~\citealt{Scholz+2020, Tendulkar+2020}). Most bursts from FRB~180916.J0158+65 have been detected within a $\sim$5.4\,d interval during cycles when the source was observed to be active (e.g., see~\citealt{Chawla+2020, CHIME+2020b, Marthi+2020, Pilia+2020, Sand+2020}), but some bursts have been found to occur slightly outside of this activity window (e.g., see~\citealt{Aggarwal+2020a}). There is now also tentative evidence for a $\sim$157\,d periodicity in the arrival times of bursts from FRB~121102, with a duty cycle of $\sim$56 percent for the activity cycle~\citep{Aggarwal+2020b, Cruces+2020, Rajwade+2020b}.

Most FRB sources have been observed at frequencies below $\sim$2\,GHz due to the smaller~FoV of radio telescopes at high frequencies, which limits the instrument's sky survey speed. As a result, the broadband spectral behavior of most FRBs remains largely unexplored at high frequencies since precise sky positions are generally needed for follow-up high frequency radio observations. The precise localization of FRB~121102 to a host galaxy~\citep{Chatterjee+2017, Marcote+2017, Tendulkar+2017} subsequently enabled the detection of numerous radio bursts up to $\sim$8\,GHz~(e.g., see~\citealt{Law+2017, Scholz+2017a, Gajjar+2018, Michilli+2018a, Spitler+2018, Zhang+2018, Gourdji+2019, Houben+2019, Pearlman+2019b, Majid+2020}). These observations revealed that FRB~121102 emits narrowband bursts, with fractional emission bandwidths of $\sim$10--30\%, across a wide range of radio frequencies. Many of these bursts also display complex time-frequency features. The emission bandwidths and sub-burst drift rates observed from FRB~121102 are typically larger at higher frequencies, on average~\citep{Gajjar+2018, Zhang+2018, Hessels+2019}. In addition, the apparent burst activity of FRB~121102 was shown to strongly depend on the range of radio frequencies that are being observed~(e.g., see~\citealt{Law+2017, Gourdji+2019, Houben+2019, Majid+2020}).

High frequency radio observations of FRBs are especially important for studying sources in the local Universe since the emission from more cosmologically distant sources will be redshifted toward lower radio frequencies~\citep{Ravi+2019b, Lu+2020b}. If the intrinsic energy distribution of bursts from FRBs is described by a steep power law (e.g., d$N$/d$E$\,$\propto$\,$E^{-\gamma}$, where $\gamma$\,$\gtrsim$\,1.8), then luminous bursts will be detected more rarely and most repeaters should be found at lower redshifts. Since the progenitor population of FRBs remains poorly constrained, broadband radio observations across a wide range of frequencies are crucial for understanding their underlying emission mechanism(s). Observations at high radio frequencies also offer a valuable resource for studying the circumburst environments of FRBs since some bursts may be rendered undetectable at lower frequencies due to a combination of free-free absorption by thermal electrons in the intervening medium, scatter-broadening produced by multipath propagation through media with electron density fluctuations, plasma lensing~\citep{Cordes+2017}, and induced Compton scattering~\citep{Ravi+2019b, Rajwade+2020a}.

In this Letter, we present results from a series of radio observations of two repeating~FRB sources, FRB~121102 and FRB~180916.J0158+65, performed simultaneously at 2.3 and 8.4\,GHz, and separately at 1.5\,GHz, with the NASA Deep Space Network~(DSN) 70-m telescopes, DSS-63 and DSS-14. The radio observations are described in Section~\ref{Section:Observations}, and the data analysis procedures and algorithms used to search for radio bursts are described in Section~\ref{Section:Analysis}. In Section~\ref{Section:Results}, we report the results of our searches for radio bursts from both of these repeating sources and list the measured properties of the bursts detected during our multiwavelength radio campaign. In Section~\ref{Section:Discussion}, we discuss the spectral properties of radio bursts from both of these repeaters and the apparent frequency dependence of the observed burst activity. We also discuss the implications of our results on the activity window and temporal distribution of radio bursts from FRB~180916.J0158+65 and place our results in the context of progenitor models proposed to explain the emission behavior of repeating~FRBs. Lastly, we provide a summary of our results and conclusions in Section~\ref{Section:Conclusions}.


\section{Radio Observations}
\label{Section:Observations}

As part of a long-term radio monitoring program of repeating FRBs with the radio telescopes comprising NASA's DSN~\citep{Pearlman+2019a}, we carried out high frequency observations of two repeating FRB sources, FRB~121102 and FRB~180916.J0158+65, using two of the DSN's large 70-m radio telescopes (DSS-63 and DSS-14). DSS-63 is located at the Madrid Deep Space Communications Complex~(MDSCC) in Robledo, Spain, and DSS-14 is located at the Goldstone Deep Space Communications Complex~(GDSCC) in Goldstone, California. Radio observations of FRB~121102 were performed between 2019~September~19 (MJD~58745) and 2020~February~11 (MJD~58890) using DSS-63. Roughly $\sim$2 weeks prior to the start of these observations, we detected 6~bursts from FRB~121102 in the 2.25\,GHz frequency band on 2019~September~6 (MJD~58732) during simultaneous 2.25 and 8.36\,GHz observations with the DSN's 70-m radio telescope (DSS-43), located at the Canberra Deep Space Communications Complex~(CDSCC) in Tidbinbilla, Australia~\citep{Majid+2020}. We used both DSS-63 and DSS-14 to observe FRB~180916.J0158+65 between 2019~September~19 (MJD~58745) and 2020~May~14 (MJD~58983). The observations of FRB~121102 and FRB~180916.J0158+65 were performed using the positions provided in~\citet{Marcote+2017} and~\citet{Marcote+2020}, respectively.

During each radio observation of FRB~121102 and FRB~180916.J0158+65 with DSS-63, we used the telescope's cryogenically cooled dual circular polarization receivers to simultaneously record $S$-band and $X$-band data at center frequencies of 2.3 and 8.4\,GHz, respectively, except during some observations where only one circular polarization channel was available at $S$-band. The system's ultra-wideband pulsar backend allowed us to simultaneously receive and save channelized power spectral densities across both frequency bands with a frequency resolution of 0.464\,MHz and time resolutions ranging between 0.28 and 2.21\,ms. The $S$-band system has a bandwidth of roughly 120\,MHz, and the $X$-band system has a bandwidth of approximately 400\,MHz. The start time, exposure time, center frequency, recorded bandwidth, number of recorded polarizations, and time resolution of each radio observation are provided in Tables~\ref{Table:Table1} and~\ref{Table:Table2}. In addition, the very long baseline interferometry~(VLBI) baseband recorder at the MDSCC was used to simultaneously record data at $S$-band and $X$-band during most observations, which allowed us to search for bursts using high time resolution data. A detailed analysis of the baseband data will be presented in an upcoming publication.

We also observed FRB~180916.J0158+65 at a center frequency of 1.5\,GHz ($L$-band) during 8~epochs using DSS-14 (see Table~\ref{Table:Table2}). While the $L$-band system on DSS-14 is capable of recording roughly 500\,MHz of total bandwidth, only 250\,MHz of the bandwidth was usable after RFI mitigation. The $L$-band data were recorded with a frequency and time resolution of 0.625\,MHz and 102.4\,$\mu$s, respectively. These $L$-band observations were also discussed in~\citet{Scholz+2020}, along with simultaneous $S$-band and $X$-band observations of FRB~180916.J0158+65 using DSS-63 during 7~separate epochs.

The data from each frequency band were flux-calibrated using the elevation-corrected system temperature, $T_{\text{sys}}$, and the radiometer equation~\citep{McLaughlin+2003}:
\begin{equation}
S_{\text{peak}}=\frac{\beta\,T_{\text{sys}}\,\text{(S/N)}_{\text{peak}}}{G\sqrt{\Delta\nu\,n_p\,t_{\text{peak}}}}.
\label{Equation:FluxDensity}
\end{equation}
Here, $\beta$\,$\approx$\,1 is a correction factor that accounts for system imperfections, such as digitization of the signal, (S/N)$_{\text{peak}}$ is the peak signal-to-noise ratio~(S/N), $G$\,$\approx$1\,K/Jy is the gain of the DSN's 70-m telescopes (DSS-63 and DSS-14), $\Delta\nu$ is the observing bandwidth, $n_p$ is the number of polarizations, and $t_{\text{peak}}$ denotes the integration time at the peak. The system temperature was measured at each frequency at the start of each observation using a noise diode modulation scheme while the antenna was pointed at zenith. The corrections applied to the $T_{\text{sys}}$ values for elevations less than 20$^{\circ}$ were minimal.


\section{Data Analysis and Searches for Radio Bursts}
\label{Section:Analysis}

The channelized filterbank data from DSS-63 and DSS-14 were processed using data reduction procedures similar to those described in previous single pulse studies of pulsars, magnetars, and FRBs with the DSN~(e.g., see~\citealt{Majid+2017, Pearlman+2018, Pearlman+2019a, Pearlman+2019b, Majid+2020, Pearlman+2020}). We first corrected the bandpass slope across the frequency band in each data set. Then, we identified frequency channels that were corrupted by radio frequency interference~(RFI) using an iterative filtering algorithm, where the time-averaged bandpass value of each frequency channel was compared to the moving median of the bandpass values. If the time-averaged bandpass value of an individual frequency channel differed from the moving median value by more than the moving standard deviation of the moving median values, then we flagged the frequency channel for masking. The moving median and moving standard deviation statistics were calculated using a sliding window of 8~frequency channels. This procedure was repeated, after replacing the bandpass value of each flagged frequency channel with its corresponding moving median value at the end of each iteration, until no additional frequency channels were flagged for removal. This algorithm identified most of the frequency channels corrupted by RFI in only a few iterations. A small number of aberrant frequency channels were also identified and masked after visually inspecting the data using the PSRCHIVE software package~\citep{Hotan+2004}. Next, in order to remove low frequency temporal variability, the moving average was subtracted from each data value in each frequency channel using a sliding window spanning 0.5\,s around each time sample.

Most of the radio bursts previously detected from FRB~121102 have observed DMs between roughly \text{500--600}\,pc\,cm$^{\text{--3}}$~(e.g., see~\citealt{Gajjar+2018, Zhang+2018, Hessels+2019, Majid+2020}). The observed~DMs of radio bursts detected thus far from FRB~180916.J0158+65 range between approximately \text{340--360}\,pc\,cm$^{\text{--3}}$~(e.g., see~\citealt{CHIME+2019b, CHIME+2020b}\footnote{See \href{https://www.chime-frb.ca/repeaters/180916.J0158+65}{https://www.chime-frb.ca/repeaters/180916.J0158+65.}\label{FootnoteCHIMEWebsite}}). Based on this, we dedispersed the cleaned filterbank data from FRB~121102 with trial DMs between 400 and 700\,pc\,cm$^{\text{--3}}$, which were linearly spaced by 5\,pc\,cm$^{\text{--3}}$ at $S$-band and 50\,pc\,cm$^{\text{--3}}$ at $X$-band. The cleaned data from FRB~180916.J0158+65 were dedispersed with trial DMs between 300 and 400\,pc\,cm$^{\text{--3}}$ using a linear DM spacing of 2\,pc\,cm$^{\text{--3}}$ at $L$-band, 5\,pc\,cm$^{\text{--3}}$ at $S$-band, and 50\,pc\,cm$^{\text{--3}}$ at $X$-band. This dedispersion scheme was chosen so that the DM smearing was less than sampling time for each observation.

A list of FRB candidates were generated using a Fourier domain matched filtering algorithm (e.g., see~\citealt{Pearlman+2018, Pearlman+2020, Majid+2020}), which was adapted from the \texttt{PRESTO} pulsar search software package\footnote{See \href{https://github.com/scottransom/presto}{https://github.com/scottransom/presto.}}~\citep{Ransom+2001}. Each full time resolution dedispersed time series was convolved with boxcar functions with logarithmically spaced widths between the native time resolution of each observation and $\sim$30.7\,ms. We recorded a list of FRB candidates with detection~S/Ns above~6.0. If a candidate was detected from the same section of data using multiple boxcar widths, only the highest S/N~event was saved in the final list. The detection~S/N of each candidate was determined using:
\begin{equation}
\text{S/N}=\frac{\sum_{i}(f_{i}-\bar{\mu})}{\bar{\sigma}\sqrt{w}},
\label{Equation:DetectionSNR}
\end{equation}
where $f_{i}$ is the time series value in bin $i$ of the boxcar function, $\bar{\mu}$ and $\bar{\sigma}$ are the local mean and root-mean-square~(RMS) noise after normalization, and $w$ is the boxcar width in number of bins. Before calculating the detection~S/N of each candidate, the time series data were detrended and normalized so that $\bar{\mu}$\,$\approx$\,0 and $\bar{\sigma}$\,$\approx$\,1. A composite list of FRB candidates was constructed by combining the candidate lists obtained from each DM trial.

We used a GPU-accelerated machine learning pipeline, which incorporates a state-of-the-art deep neural network from the \texttt{FETCH}\footnote{See \href{https://github.com/devanshkv/fetch}{https://github.com/devanshkv/fetch.}} (Fast Extragalactic Transient Candidate Hunter) software package~\citep{Agarwal+2020}, to identify astrophysical bursts from among the large sample of FRB candidates returned by the Fourier domain matched filtering algorithm. Probabilities~($p$) were assigned to each candidate using the \texttt{DenseNet121} Frequency-Time (FT)/\texttt{Xception} DM-Time~(DMT) ``a'' model (see Table~4 in~\citealt{Agarwal+2020}), trained using a transfer learning approach. The probability associated with each candidate indicated the likelihood that the candidate was astrophysical. Diagnostic plots of all candidates with $p$\,$>$\,0.3 were visually inspected for verification.


\section{Results}
\label{Section:Results}

\subsection{FRB~121102}
\label{Section:FRB121102}

\begin{figure*}[b]
	\centering
	\begin{tabular}{cc}
		
		\subfigure
		{
			\includegraphics[trim=0cm 0cm 0cm 0cm, clip=false, scale=0.37, angle=0]{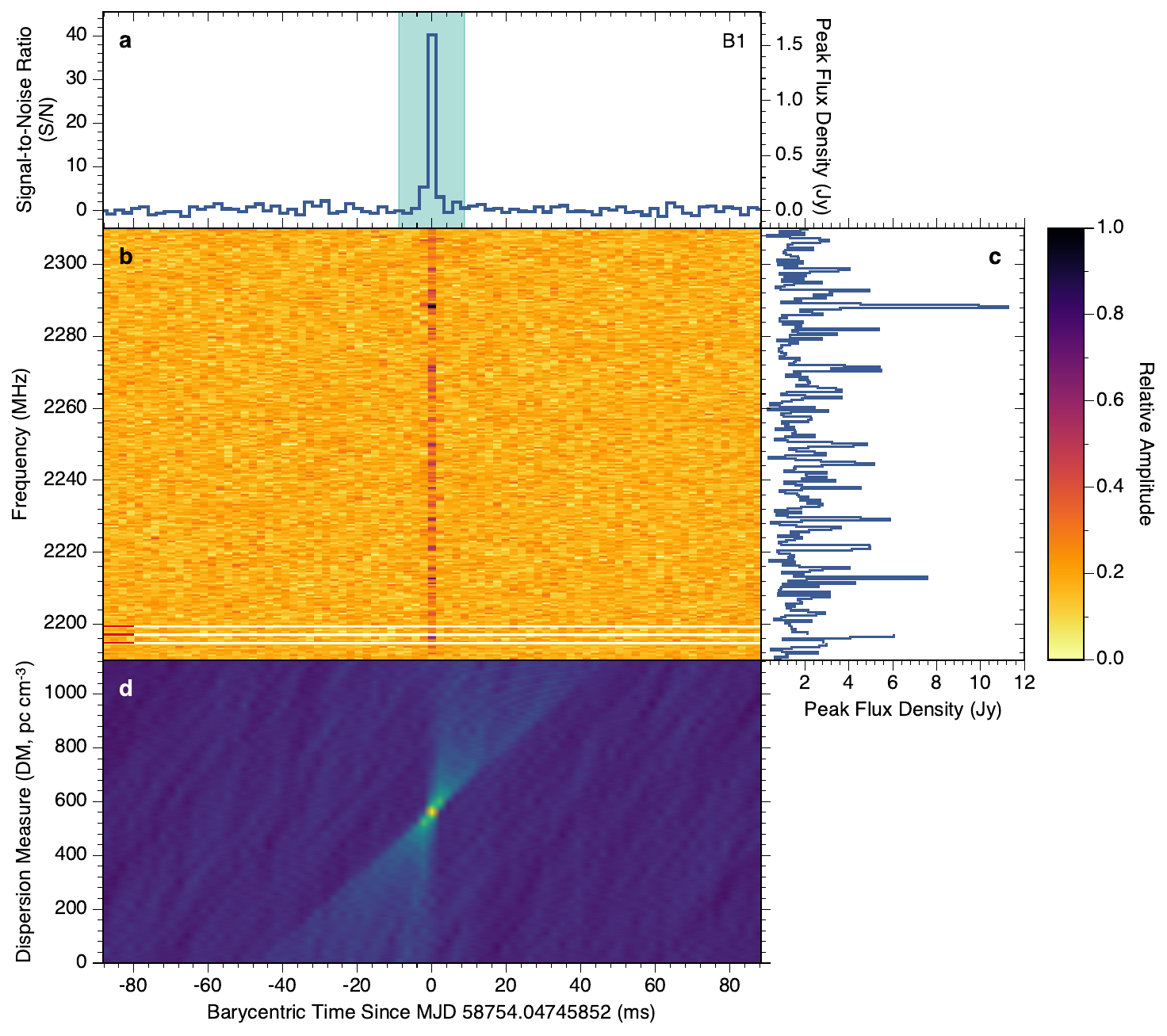}
			\label{Figure:Figure1a}
		}
		
		&
		
		\subfigure
		{
			\includegraphics[trim=0cm 0cm 0cm 0cm, clip=false, scale=0.37, angle=0]{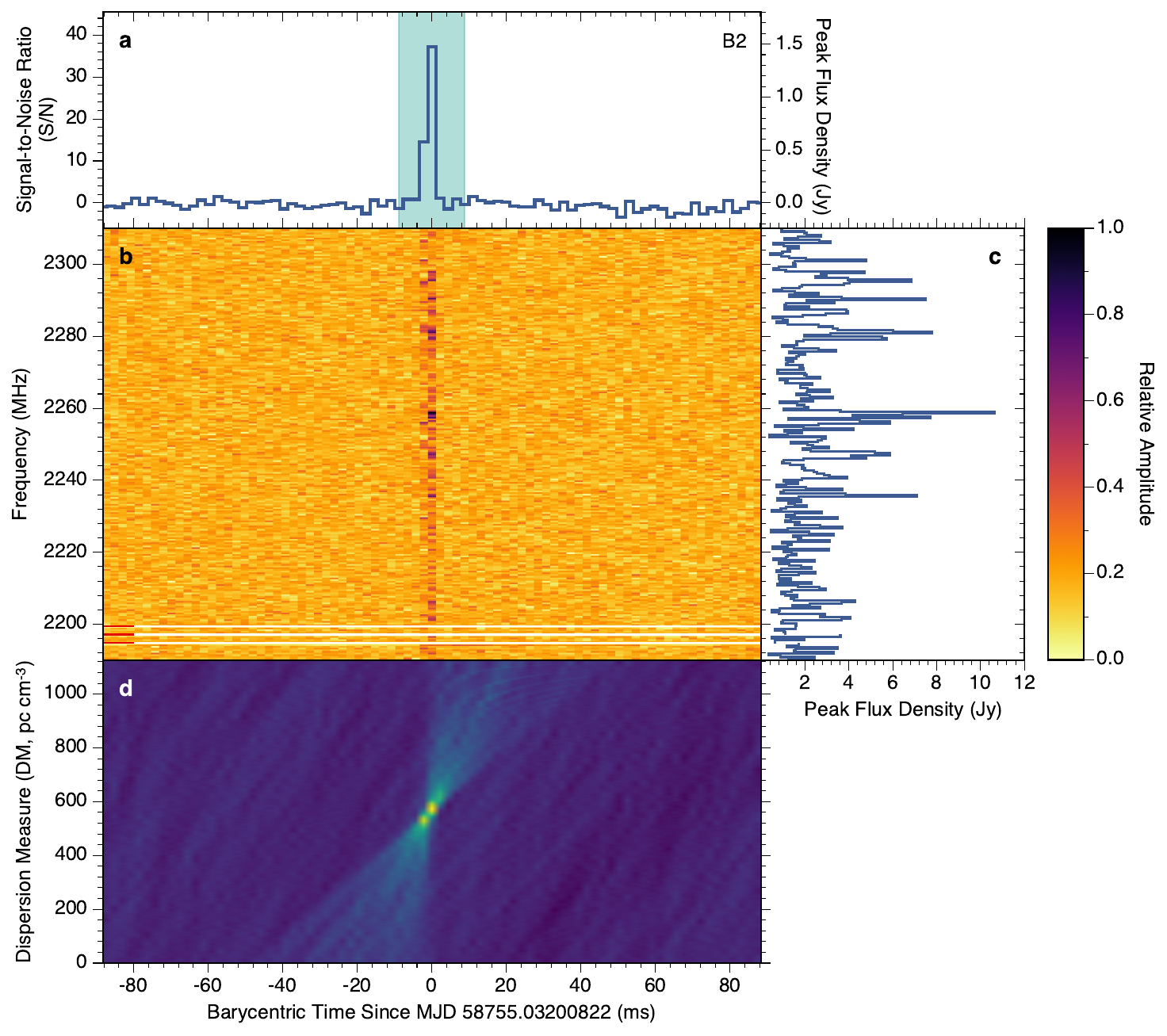}
			\label{Figure:Figure1b}
		}
		
	\end{tabular}
	\caption{$S$-band radio bursts (B1 and B2) detected from FRB~121102 on (left)~2019~September~28 (MJD~58754) and (right)~2019~September~29 (MJD~58755) using DSS-63. The frequency-averaged burst profiles are shown in panel~(a), and the dedispersed dynamic spectra are displayed in panel~(b). The data are shown with a time and frequency resolution of 2.2\,ms and 0.464\,MHz, respectively. The color bar on the right shows the relative amplitude of the burst's spectral-temporal features. The solid white lines and red markers in panel~(b) indicate frequency channels that have been masked due to the presence of radio frequency interference~(RFI). The flux-calibrated burst spectra are shown in panel~(c). The teal shaded region in panel (a) corresponds to the interval used for extraction of the on-pulse spectrum in panel (c). Both bursts have been dedispersed using a dispersion measure~(DM) of 563.0\,pc\,cm$^{\text{--3}}$, which corresponds to the average DM near the time of each burst (A. D. Seymour, private communication). The DM--time images of each burst are displayed in panel~(d) and show the signal-to-noise ratio~(S/N) of each burst after dedispersion.}
	\label{Figure:Figure1}
\end{figure*}

FRB~121102 was observed simultaneously at $S$-band and $X$-band with DSS-63 for 27.3\,hr between 2019~September~19 (MJD~58745) and 2020~February~11 (MJD~58890) during a recent period of activity from the source. We detected 2~bursts at $S$-band during these observations. The first burst~(B1) was detected on 2019~September~28 (MJD~58754), and the second burst~(B2) was detected approximately one day later on 2019~September~29 (MJD~58755). In Figure~\ref{Figure:Figure1}, we show the frequency-averaged profiles, dedispersed dynamic spectra, flux-calibrated burst spectra, and DM-time images of each burst. Both of these bursts were detected in filterbank data recorded with a time resolution of 2.2\,ms, and thus they are not temporally resolved. The apparent~DM associated with a particular radio burst from a given FRB source can differ depending on the spectral-temporal structure of the burst and whether a signal-to-noise-maximizing or structure-maximizing metric is used to determine the optimal DM. Since we were unable to resolve the spectral-temporal structure of the bursts (B1 and B2) due to the time resolution of the data, we dedispersed both bursts in Figure~\ref{Figure:Figure1} using a DM of 563.0\,pc\,cm$^{\text{--3}}$, which corresponds to the average~DM near the time of each burst obtained from long-term monitoring of FRB~121102 with the Arecibo Observatory~(A. D. Seymour, private communication).

In Table~\ref{Table:Table3}, we provide a list of measured properties of each burst, including the barycentric arrival time, peak~S/N, burst width, peak flux density, time-integrated burst fluence~($\mathcal{F}$), spectral energy density, and isotropic-equivalent energy. The burst widths were determined by fitting a Gaussian function to the dedispersed burst profile. We quote the full width at half maximum~(FWHM) as the temporal width. The burst fluence was determined using the 2$\sigma$~FWHM for the duration of each burst.

There was no evidence of radio emission at $X$-band during the times when bursts were detected at $S$-band. Baseband data was not available at either frequency band during the two epochs when the $S$-band bursts were detected. We also did not detect any $X$-band bursts from FRB~121102 at any other times during our observations, despite the fact that the $X$-band bandwidth was a factor of $\sim$3.5 larger than the bandwidth at $S$-band and the 6$\sigma$~fluence detection thresholds were roughly two times lower at $X$-band than at $S$-band, on average. In Table~\ref{Table:Table1}, we list the 6$\sigma$~fluence detection thresholds ($\mathcal{F_\text{min}}$) for each observation and each frequency band.


\subsection{FRB~180916.J0158+65}
\label{Section:FRB180916}

During $\sim$90.8\,hr of simultaneous $S$-band and $X$-band observations of FRB~180916.J0158+65, carried out between 2019~September~19 (MJD~58745) and 2020~May~14 (MJD~58983) with DSS-63, no radio bursts were detected at either frequency band. We also observed FRB~180916.J0158+65 at $L$-band with DSS-14 for $\sim$2.1\,hr on 2019~December~2 (MJD~58819) and for $\sim$8.9\,hr on 2019~December~18 (MJD~58835), but did not detect any radio bursts (see also~\citealt{Scholz+2020}). The 6$\sigma$~fluence detection thresholds ($\mathcal{F_\text{min}}$) associated with each observation and frequency band are listed in Table~\ref{Table:Table2}.

\begin{figure*}[b]
	\centering
	\hspace{-1.28cm} 
	\includegraphics[trim=0cm 0cm 0cm 0cm, clip=false, scale=0.42, angle=0]{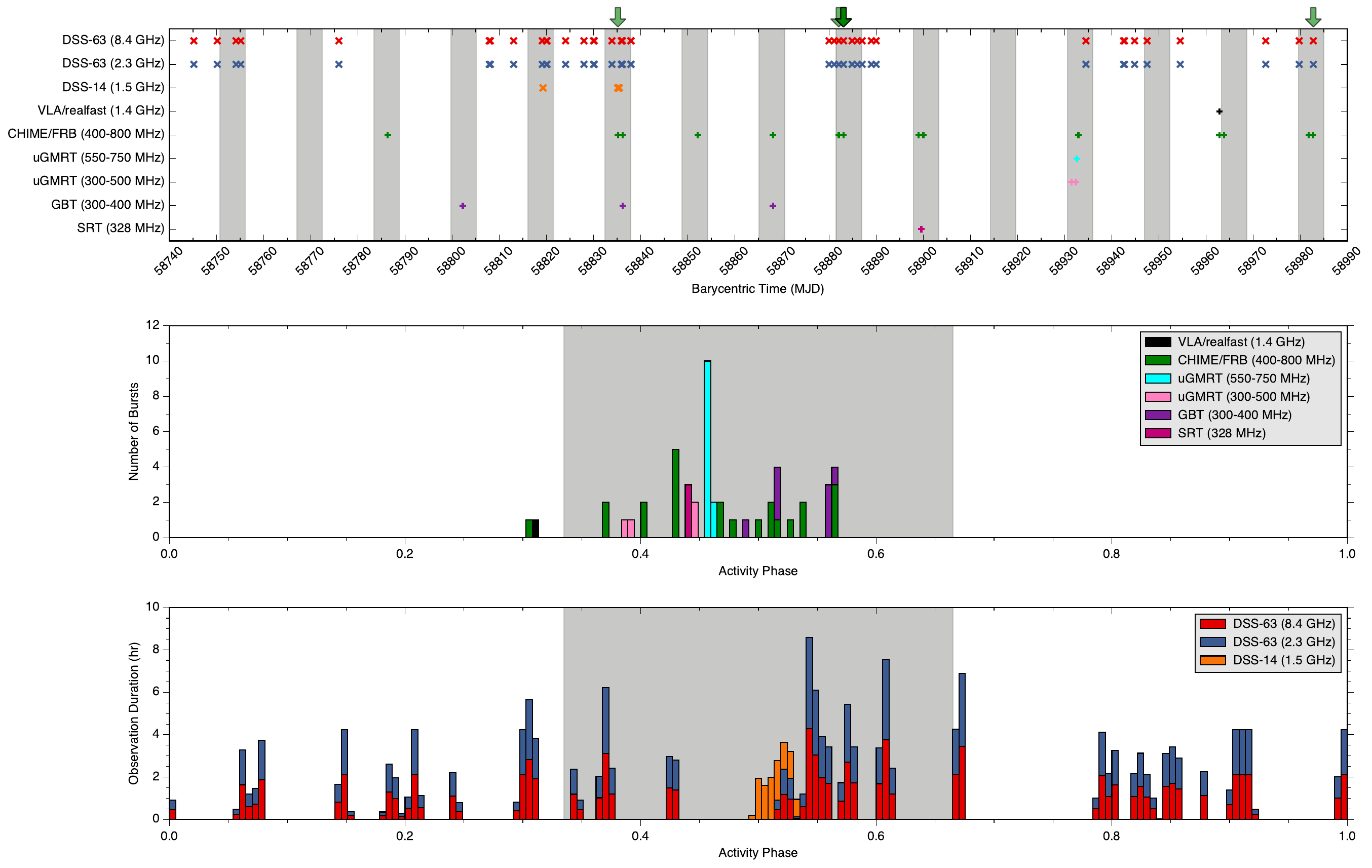}
	\caption{Top panel: Radio observations of FRB~180916.J0158+65 between 2019~September~19 (MJD~58745) and 2020~May~14 (MJD~58983). The barycentric time at the middle of the simultaneous 8.4 and 2.3\,GHz observations with DSS-63 are labeled using red and blue crosses ($\times$), respectively. The barycentric mid-time of the 1.5\,GHz observations with DSS-14 are indicated by orange crosses. The barycentric times of the observations with DSS-63 and DSS-14 were determined by removing the time delay from dispersion using a dispersion measure~(DM) of 348.82\,pc\,cm$^{\text{--3}}$ and correcting to infinite frequency using the position ($\alpha_{\text{J2000}}$\,$=$\,01$^{\text{h}}$58$^{\text{m}}$00$^{\text{s}}$.75017, $\delta_{\text{J2000}}$\,$=$\,65$^{\circ}$43$\arcmin$00$\arcsec$.3152) in \citet{Marcote+2020}. The duration of each of the DSN observations (see Table~\ref{Table:Table2}) is shown using horizontal error bars, which are smaller than the symbol size. The barycentric times of bursts detected from FRB~180916.J0158+65 with the Very Large Array~(VLA)/realfast at 1.4\,GHz~\citep{Aggarwal+2020b}, CHIME/FRB radio telescope between 400--800\,MHz~\citep{CHIME+2020b}\textsuperscript{\ref{FootnoteCHIMEWebsite}}, upgraded Giant Metrewave Radio Telescope~(uGMRT) between 550--750 and 300--500\,MHz~\citep{Marthi+2020, Sand+2020}, Robert C. Byrd Green Bank Telescope~(GBT) between 300--400\,MHz~\citep{Chawla+2020}, and the Sardinia Radio Telescope~(SRT) at 328\,MHz~\citep{Pilia+2020} are labeled using black, green, cyan, pink, purple, and magneta pluses ($+$), respectively. In total, 8 of the radio bursts detected by CHIME/FRB occurred during times when DSS-63/DSS-14 was also observing FRB~180916.J0158+65, and these bursts are labeled using green arrows and listed in Table~\ref{Table:Table4}. The gray shaded regions correspond to a $\pm$2.7\,d window around the peak of FRB~180916.J0158+65's activity phase, assuming an activity period of 16.35\,d~\citep{CHIME+2020b}. Middle panel: Histogram of activity phases at which each of the radio bursts in the top panel were detected. The activity phase of the bursts are shown using colored bars, where each instrument is labeled using the color used in the top panel. Bottom: Exposure times of radio observations of FRB~180916.J0158+65 using DSS-63 and DSS-14 between 2019~September~19 and 2020~May~14 as a function of FRB~180916.J0158+65's activity phase. The exposure times of the 8.4\,GHz observations with DSS-63, 2.3\,GHz observations with DSS-63, and 1.5\,GHz observations with DSS-14 are indicated using red, blue, and orange bars, respectively. In both the middle and bottom panels, the width of each bar is $\sim$0.0054 phase units, which is equal to the average duration ($\sim$2.1\,hr) of the DSN~radio observations. The gray shaded regions, in the middle and bottom panels, correspond to a $\pm$2.7\,d window around the peak of FRB~180916.J0158+65's activity phase.}
	\label{Figure:Figure2}
\end{figure*}

In the top panel of Figure~\ref{Figure:Figure2}, we show the barycentric mid-time of each of our radio observations of FRB~180916.J0158+65 with DSS-63 and DSS-14 between 2019~September~19 and 2020~May~14, after removing the time delay due to dispersion using a~DM of 348.82\,pc\,cm$^{\text{--3}}$ and correcting to infinite frequency. This~DM corresponds to the average structure-optimizing DM derived from four bursts detected with the CHIME/FRB baseband system~\citep{CHIME+2020b}. We also show the barycentric times of radio bursts detected from FRB~180916.J0158+65 using the Very Large Array~(VLA)/realfast at 1.4\,GHz~\citep{Aggarwal+2020b}, CHIME/FRB instrument between 400--800\,MHz~\citep{CHIME+2020b}\textsuperscript{\ref{FootnoteCHIMEWebsite}}, upgraded Giant Metrewave Radio Telescope~(uGMRT) between 550--750 and 300--500\,MHz~\citep{Marthi+2020, Sand+2020}, Robert C. Byrd Green Bank Telescope~(GBT) between 300--400\,MHz~\citep{Chawla+2020}, and the Sardinia Radio Telescope~(SRT) at 328\,MHz~\citep{Pilia+2020} during this time period. A total of 8~bursts were detected by CHIME/FRB during times when we were simultaneously observing the source with DSS-63/DSS-14~\citep{CHIME+2020b}\textsuperscript{\ref{FootnoteCHIMEWebsite}}, which are labeled using green arrows in the top panel of Figure~\ref{Figure:Figure2} (also see Table~\ref{Table:Table4}). The gray shaded regions correspond to the $\pm$2.7\,d activity window around the peak of FRB~180916.J0158+65's activity phase, based on the 16.35\,d activity period~\citep{CHIME+2020b}. The bottom panel of Figure~\ref{Figure:Figure2} shows the total exposure time of the 1.5, 2.3, and 8.4\,GHz DSN radio observations as a function of FRB~180916.J0158+65's activity phase.

The barycentric arrival times and properties of the 8~radio bursts detected from FRB~180916.J0158+65 by CHIME/FRB during our DSN observing epochs are listed in Table~\ref{Table:Table4}. The C1 burst was detected by CHIME/FRB during an overlapping $L$-band observation of FRB~180916.J0158+65 with DSS-14~(see also~\citealt{Scholz+2020}), and the other 7~bursts (C2--C8) occurred during epochs when we were simultaneously observing the source at $S$-band and $X$-band with DSS-63. We extracted radio data from DSS-14/DSS-63, recorded using the pulsar backend, around each of these times and visually inspected the frequency-averaged profiles and dedispersed dynamic spectra. We found no evidence of radio emission in the pulsar backend data during any of these times.

\begin{figure*}[b]
	\centering
	\includegraphics[trim=0cm 0cm 0cm 0cm, clip=false, scale=0.39, angle=0]{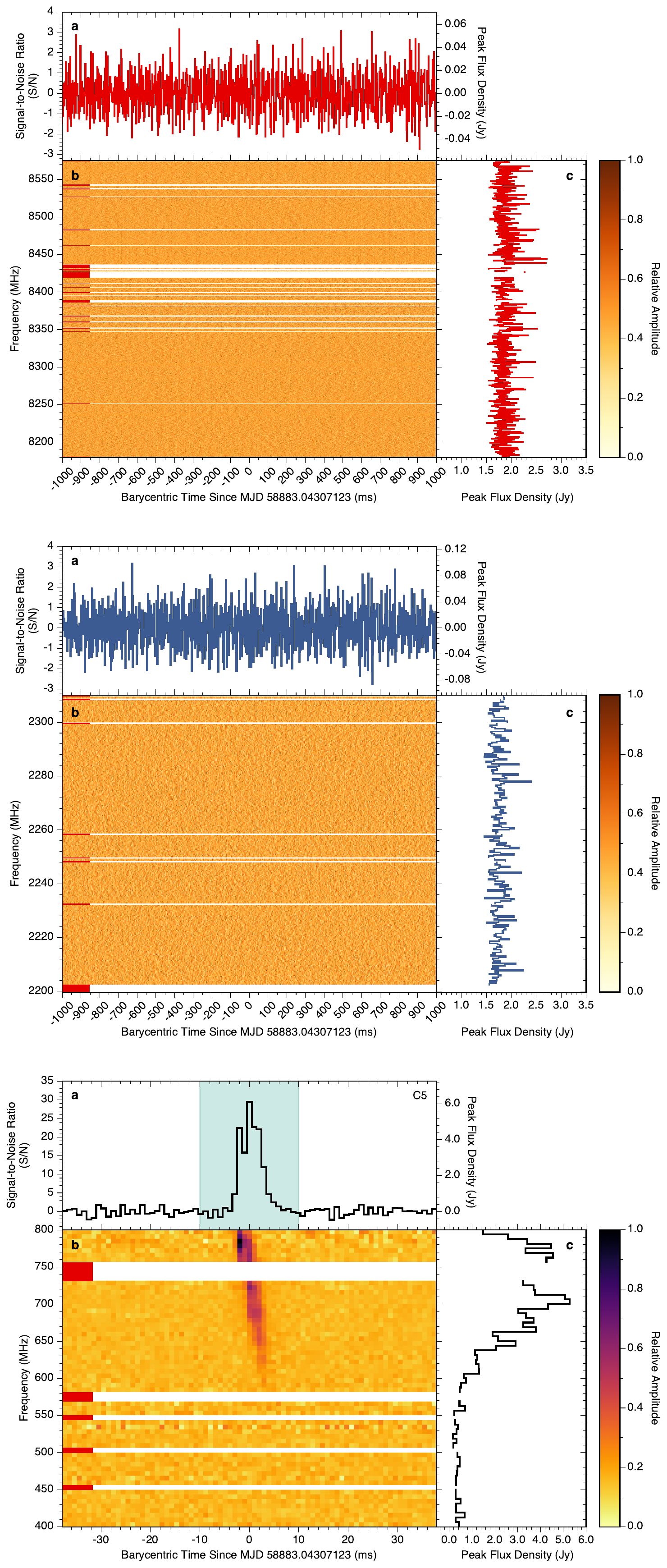}
	\caption{Simultaneous $X$-band~(top) and $S$-band~(middle) observations of FRB~180916.J0158+65 with DSS-63 during the time of a bright burst~(C5, see Table~\ref{Table:Table4}) detected by CHIME/FRB on MJD~58883.04307123, shown in the bottom panels. The data in each frequency band have been dedispersed using a dispersion measure~(DM) of 348.82\,pc\,cm$^{\text{--3}}$, which corresponds to the average structure-optimizing DM derived from four bursts detected with the CHIME/FRB baseband system~\citep{CHIME+2020b}. The frequency-averaged Stokes~I profiles are shown in panel (a), and the dedispersed Stokes~I dynamic spectra are displayed in panel~(b). The $X$-band and $S$-band data are shown with a time and frequency resolution of 2.2\,ms and 0.464\,MHz, respectively. The CHIME/FRB data are shown with a time resolution of 0.98304\,ms, and we have downsampled the full-resolution data (16,384 channels) into 64 sub-bands, which each have a bandwidth of 6.25\,MHz, for better visualization. The color bar on the right shows the relative amplitude of the spectral-temporal features in the dedispersed dynamic spectra. Frequency channels that have been masked due to the presence of radio frequency interference~(RFI) are indicated using solid white lines and red markers in panel~(b). The flux-calibrated spectra are shown in panel~(c). For the burst detected by CHIME/FRB, the teal shaded region in panel (a) corresponds to the interval used to extract the on-pulse spectrum in panel (c).}
	\label{Figure:Figure3}
\end{figure*}

The C5~burst reported by~\citet{CHIME+2020b} on MJD~58883.04307123 is one of the highest fluence bursts ($\mathcal{F}$\,$=$\,16.3\,$\pm$\,5.0\,Jy\,ms) detected from FRB~180916.J0158+65 in the 400--800\,MHz band thus far. In Figure~\ref{Figure:Figure3}, we show the frequency-averaged profiles, dedispersed dynamic spectra, and flux-calibrated spectra, spanning $\pm$1\,s around the time of the burst, during our simultaneous $S$-band and $X$-band observations of FRB~180916.J0158+65 with DSS-63 using the pulsar backend recorder, along with the C5 burst detected by CHIME/FRB. We also extracted baseband data at $S$-band and $X$-band around the time of this burst to rule out the presence of faint, narrow-width radio bursts that may have been missed in the pulsar backend data. After inspecting the high time resolution data, no evidence of radio emission was found at either frequency band. A detailed analysis of the baseband data obtained during the epochs listed in Table~\ref{Table:Table2} will be presented in an upcoming publication.


\section{Discussion}
\label{Section:Discussion}

\subsection{Spectral Properties of Radio Bursts}
\label{Section:RadioSpectrum}

Multifrequency, broadband radio observations of repeating~FRBs have demonstrated that their burst spectra are highly variable and often peak in a narrow frequency band. Some bursts display emission that is limited to a frequency range of tens to hundreds of MHz~(e.g., see~\citealt{Gourdji+2019, CHIME+2019b, Hessels+2019, CHIME+2020b, Kumar+2020, Majid+2020}), while others appear to have a frequency extent that spans at least $\sim$2\,GHz~(e.g., see~\citealt{Law+2017}). Many bursts from repeating~FRBs also show evidence of downward-drifting subpulses~\citep{CHIME+2019a, CHIME+2019b, Hessels+2019, Josephy+2019}. On average, larger sub-burst bandwidths and drift rates have been observed from repeating~FRBs at higher radio frequencies~\citep{Gourdji+2019, Hessels+2019, Josephy+2019, Chawla+2020}. In many cases, the spectra of FRBs from repeating sources, such as FRB~121102, are not well-modeled by a power law with a single spectral index~(e.g., see~\citealt{Scholz+2016, Spitler+2016, Law+2017}). The assortment of spectral behavior may be intrinsic to the source's underlying emission mechanism(s), produced by propagation effects, or generated by a combination of radiation and propagation processes.

While radio bursts from FRB~121102 have been detected up to a frequency of $\sim$8\,GHz~\citep{Gajjar+2018}, no FRB source has yet been observed to produce emission at higher radio frequencies. At 1.4\,GHz, bursts from FRB~121102 have characteristic bandwidths of $\sim$250\,MHz, with a 1$\sigma$ variation of $\sim$90\,MHz~\citep{Hessels+2019}. The characteristic bandwidths measured from bursts detected from FRB~121102 between 4.5--8\,GHz with the GBT range from $\sim$1--100\,MHz~\citep{Gajjar+2018}. However, these bursts also display large-scale frequency structure, with a frequency extent of $\sim$1\,GHz, in the 4.5--8\,GHz frequency band.

The Galactic pulse broadening timescale predicted along the line of sight to FRB~121102 by the NE2001 Galactic electron density model~\citep{Cordes+2002} is:
\begin{equation}
\tau_{d,\text{121102}}\,=\,\text{22}\Big(\frac{\nu}{\text{1\,GHz}}\Big)^{-\alpha}\,\mu\text{s},
\label{Equation:PulseBroadeningTimescaleR1}
\end{equation}
where $\nu$ is the observing frequency in GHz and $\alpha$ is the pulse broadening spectral index. The expected diffractive interstellar scintillation~(DISS) bandwidth ($\Delta\nu_{\text{DISS}}$) is given by~\citep{Cordes+1998}:
\begin{equation}
\Delta\nu_{\text{DISS}}=\frac{C_{\text{1}}}{\text{2}\pi\tau_{d}}=\text{0.0084}\Big(\frac{\nu}{\text{1\,GHz}}\Big)^{\alpha}\,\text{MHz},
\label{Equation:DISSBandwidth}
\end{equation}
where $C_{\text{1}}$\,$=$\,1.16 for a uniform, Kolomogorov medium, and we have substituted $\tau_{d,\text{121102}}$ in Equation~\eqref{Equation:PulseBroadeningTimescaleR1} for $\tau_{d}$ in Equation~\eqref{Equation:DISSBandwidth}. In general, the precise value of $C_{\text{1}}$ depends on both the geometry and wavenumber spectrum of the electron density. Assuming a pulse broadening spectral index between 4 and 4.4, the DISS bandwidth, $\Delta\nu_{\text{DISS}}$, can range from \mbox{$\sim$0.01--133\,MHz} for observing frequencies between 1~and 9\,GHz. These estimates are consistent with previous measurements of the scintillation bandwidth of FRB~121102 in this frequency range~(e.g., see~\citealt{Hessels+2019, Majid+2020}). Many bursts from FRB~121102 also display spectral features with a frequency extent that cannot be attributed solely to Galactic DISS.

Radio observations of repeating~FRBs over a wide frequency range have revealed bright bursts at lower frequencies that do not have detectable radio emission at higher frequencies. For example, \citet{Majid+2020} carried out simultaneous radio observations of FRB~121102 at 2.25\,GHz ($S$-band) and 8.36\,GHz ($X$-band) with the 70-m DSN radio telescope, DSS-43, and discovered 6~bursts in the 2.25\,GHz frequency band. No radio emission was detected in the 8.36\,GHz frequency band at the same time, despite having greater sensitivity and a larger bandwidth at $X$-band. Both of the $S$-band bursts (B1 and B2), shown in Figure~\ref{Figure:Figure1}, from FRB~121102 also display prominent emission at 2.3\,GHz without accompanying radio emission at 8.4\,GHz at the same time. The spectral behavior observed in these frequency bands, which are separated by two octaves in frequency, cannot be attributed to Galactic scintillation. Our observations of FRB~121102 further demonstrate that the source's apparent burst activity strongly depends on the radio frequency band that is being utilized.

FRB~180916.J0158+65 is one of the most prolifically bursting repeating FRB sources known and displays a significant 16.35\,d periodicity in the arrival times of its bursts~\citep{CHIME+2020b}. Multiple radio bursts have been detected from FRB~180916.J0158+65 thus far between $\sim$300\,MHz and $\sim$1.7\,GHz using various radio telescopes (e.g., see~\citealt{CHIME+2019b, Aggarwal+2020a, Chawla+2020, CHIME+2020b, Marcote+2020, Marthi+2020, Pilia+2020, Sand+2020, Scholz+2020}). It is surprising that no bursts were detected from FRB~180916.J0158+65 during our 101.8\,hr monitoring campaign at 1.5, 2.3, and 8.4\,GHz with DSS-14 and DSS-63, considering the high level of activity observed from the source at lower frequencies (e.g., see~\citealt{CHIME+2020b}). Our observations of FRB~180916.J0158+65 further demonstrate that the emission process is strongly frequency-dependent~(e.g., see~\citealt{Chawla+2020, CHIME+2020b}), and its bursts are band-limited, with less spectral occupancy at high frequencies ($\gtrsim$\,2\,GHz). These observations also suggest that the source emits fewer or fainter bursts in the 2.3 and 8.4\,GHz frequency bands compared to the emission behavior observed at lower radio frequencies. A detailed discussion of the source's burst rates is provided in Section~\ref{Section:TemporalDistribution}.

Toward the direction of FRB~180916.J0158+65, the Galactic pulse broadening timescale predicted by the NE2001 Galactic electron density model~\citep{Cordes+2002} is:
\begin{equation}
\tau_{d,\text{180916}}\,=\,\text{21}\,\mu s\Big(\frac{\nu}{\text{1\,GHz}}\Big)^{-\alpha},
\label{Equation:PulseBroadeningTimescaleR3}
\end{equation}
which is comparable to the estimate obtained from FRB~121102 in Equation~\eqref{Equation:PulseBroadeningTimescaleR1}. Substituting $\tau_{d,\text{180916}}$ for $\tau_{d}$ in Equation~\eqref{Equation:DISSBandwidth} and assuming a pulse broadening spectral index between 4 and 4.4, we find that the DISS bandwidth can range from $\sim$0.01--137\,MHz for observing frequencies between 1 and 9\,GHz, which is consistent with the 0.06\,MHz scintillation bandwidth measured from the brightest burst detected at 1.7\,GHz with the EVN~\citep{Marcote+2020}. Therefore, during the times that bursts were detected from FRB~180916.J0158+65 by CHIME/FRB, the absence of radio emission during our simultaneous observations of the source in the 2.3 and 8.4\,GHz frequency bands cannot be explained by Galactic scintillation. Instead, this behavior is likely an intrinsic property of the emission. FRB~180916.J0158+65's low rotation measure~(RM) and lack of a luminous, persistent radio source both suggest that its circumburst environment may be less extreme than that of FRB~121102, the only other repeating~FRB source localized using~VLBI~\citep{Marcote+2017, Marcote+2020}. However, both FRB~121102 and FRB~180916.J0158+65 show similar spectral behavior, which supports the notion that the behavior is instrinsic and not produced by propagation effects. While we find it unlikely that extrinsic effects in FRB~180916.J0158+65's local environment can fully account for the spectral behavior observed at high radio frequencies, they may contribute to some of the characteristic features observed across its radio spectrum. These observations also demonstrate that FRB~180916.J0158+65's apparent burst activity, like FRB~121102, strongly depends on the choice of radio observing frequency.

\subsection{Correlation Between Radio Frequency and \\ Activity Phase of Radio Bursts}
\label{Section:FrequencyActivityPhase}

After folding the radio burst arrival times from FRB~180916.J0158+65 modulo the 16.35\,d activity period, multiple bursts detected at 1.4 and 1.7\,GHz were found to occur near the leading edge of the activity cycle~\citep{Aggarwal+2020a, CHIME+2020b}. This has led to the suggestion that there may be a correlation between the emission frequency and the source's activity phase (e.g., see~\citealt{Aggarwal+2020b}), where bursts emitted at higher radio frequencies occur earlier in the activity cycle. However, previous high radio frequency observations of FRB~180916.J0158+65 have not monitored the source across a wide range of activity phases, which is necessary for determining if such a correlation exists.

The cadence of our high frequency radio observations with DSS-63 and DSS-14 across FRB~180916.J0158+65's activity phase is shown in the bottom panel of Figure~\ref{Figure:Figure2}. These observations cover a wide range of activity phases, both within the $\pm$2.7\,d activity window reported by~\citet{CHIME+2020b} and at phases outside of this range. The DSN observations also cover several activity cycles during which radio bursts were detected at lower frequencies by CHIME/FRB, indicating that the source was active during these epochs. Although our radio observations cover most of the activity phases at which bursts were detected by other instruments during our monitoring campaign, our observing exposures are not uniform across all activity phases, and there are some gaps in the phase coverage. Since no bursts were detected from FRB~180916.J0158+65 in the 1.5, 2.3, or 8.4\,GHz frequency bands during our observations with DSS-63 and DSS-14, we were unable to find evidence to support the notion that bursts emitted from the source at higher radio frequencies occur earlier in the activity cycle. However, we cannot presently exclude the possibility that a frequency-phase correlation may exist. Additional high frequency radio observations of FRB~180916.J0158+65, and the detection of bursts above $\sim$2\,GHz, will be able to further address whether or not there is a link between the emission frequency of the bursts and the activity phase at which they are detected.

\subsection{Temporal Distribution of Radio Bursts}
\label{Section:TemporalDistribution}

The discovery of a 16.35\,d activity period from FRB~180916.J0158+65, together with an activity window during which most of its radio bursts are detected, has enabled dedicated multiwavelength follow-up campaigns to quantify the activity of the source at various wavelengths (e.g., see~\citealt{Aggarwal+2020a, Chawla+2020, CHIME+2020b, Marcote+2020, Marthi+2020, Pilia+2020, Sand+2020, Scholz+2020}). We demonstrate here that the apparent burst rate from FRB~180916.J0158+65 is a strong function of radio observing frequency and discuss the temporal distribution of bursts at high radio frequencies.

Assuming Poisson statistics, the probability of randomly detecting exactly $n$ bursts during a time interval, $t$, is given by:
\begin{equation}
p(n, \lambda) = \frac{\lambda^{n}e^{-\lambda}}{n!},
\label{Equation:PoissonProbability}
\end{equation}
where $\lambda$ is the number of bursts that are expected to be detected during this time interval.
For a fixed rate parameter, $\lambda$, the probability of detecting at least $N$ bursts during a time interval, $t$, can be computed using:
\begin{equation}
P(n\ge N)=e^{-\lambda}\sum_{k=N}^{\infty}\frac{\lambda^{k}}{k!}=1-e^{-\lambda}\sum_{k=0}^{N-1}\frac{\lambda^{k}}{k!}.
\label{Equation:ProbabilityAtLeastNBursts}
\end{equation}
From Equation~\eqref{Equation:PoissonProbability}, if the probability of detecting exactly 0~bursts, $p_{0}$\,$=$\,$p(n$\,$=$\,0$,\lambda)$, is known, then the rate parameter, $\lambda$, can be calculated analytically using:
\begin{equation}
\lambda=-\ln(p_{0}).
\label{Equation:PoissonRateParameter}
\end{equation}

We observed FRB~180916.J0158+65 simultaneously at \text{$S$-band} and $X$-band with DSS-63 for a total of $t$\,$=$\,36.2\,hr during the $\pm$2.7\,d interval around the peak of the activity window. In these frequency bands, we assume that the source's emission behavior during the activity window follows Poisson statistics and that there was a non-detection probability of 95\%. Using Equation~\eqref{Equation:PoissonRateParameter}, we find that the expected number of bursts during this time interval is $\sim$0.051 bursts, which corresponds to a rate of $r$\,$\approx$\,0.0014 bursts per hour. If the above assumptions hold, then this implies that $\sim$700\,hr of radio observations would be needed to detect a single burst in these frequency bands. However, it is possible that the emission process produces bursts with waiting times that may not follow a Poisson distribution~\citep{Cruces+2020}.

The average 6$\sigma$~fluence detection thresholds during our $L$-band, $S$-band, and $X$-band observations of FRB~180916.J0158+65 with DSS-14 and DSS-63 were 0.29, 0.26, and 0.14\,Jy\,ms, respectively. We note that 3~of the 4~radio bursts detected at 1.7\,GHz from FRB~180916.J0158+65 using the European VLBI~Network~(EVN) had measured burst widths of $\sim$1.7\,ms and fluences exceeding $\sim$0.6\,Jy\,ms~\citep{Marcote+2020}, which are some of the faintest bursts detected to date. If a similarly bright burst, with a comparable temporal width and fluence, was emitted from the source at 1.5, 2.3, or 8.4\,GHz during our radio observations, it would have been detected with S/N\,$>$\,10.

In the 400--800\,MHz frequency band, radio bursts detected from FRB~180916.J0158+65 appear to be clustered in time, and the burst activity is not constant at all activity phases~\citep{CHIME+2019b, CHIME+2020b}. Assuming that the radio bursts obey Poisson statistics and can be treated as independent events, \citet{CHIME+2020b} estimated a detection rate of $r$\,$=$\,0.9$^{\text{+0.5}}_{\text{--0.4}}$ bursts per hour above a fluence threshold of 5.2\,Jy\,ms in a $\pm$2.7\,d interval around the peak of the activity window, which corresponds to activity phases between $\sim$0.33--0.67. They also estimated detection rates above this fluence threshold in 3~subintervals of the $\pm$2.7\,d activity window: (subinterval~1) $\pm$0.9\,d around the activity peak, i.e. activity phases $\sim$0.44--0.56; (subinterval~2) between 0.9 and 1.8\,d from the activity peak, i.e. activity phases $\sim$0.39--0.44 and $\sim$0.56--0.61; (subinterval~3) between 1.8 and 2.7\,d from the activity peak, i.e., activity phases $\sim$0.33--0.39 and $\sim$0.61--0.67. The estimated detection rates in subintervals~1--3 were $r_{\text{1}}$\,$=$\,1.8$^{\text{+1.3}}_{\text{--0.8}}$, $r_{\text{2}}$\,$=$\,0.8$^{\text{+1.0}}_{\text{--0.5}}$, and $r_{\text{3}}$\,$=$\,0.1$^{\text{+0.6}}_{\text{--0.1}}$ bursts per hour above a fluence of 5.2\,Jy\,ms, respectively~\citep{CHIME+2020b}. The uncertainties associated with these detection rates correspond to 95\% confidence limits.

We can also use the formalism described in~\citet{Houben+2019} to calculate the expected burst rates at 1.5, 2.3, and 8.4\,GHz with DSS-14 and DSS-63 during the activity intervals defined above. We assume that the differential rate distributions at two frequencies are related by a power law parameterized by a statistical spectral index, $\alpha_{\text{s}}$. The statistical spectral index characterizes the spectrum of the overall burst rate, which is distinct from the instantaneous spectral index of individual bursts. Here, we use $\alpha_{\text{s}}$\,$=$\,\text{--1.6}$^{\text{+1.0}}_{\text{--0.6}}$ for the statistical spectral index of FRB~180916.J0158+65~\citep{Chawla+2020}. The detection rates, $r_{\nu_{\text{1}}}$ and $r_{\nu_{\text{2}}}$, at two frequencies, $\nu_{\text{1}}$ and $\nu_{\text{2}}$, can be determined using:
\begin{equation}
\frac{r_{\nu_{\text{1}}}}{r_{\nu_{\text{2}}}}=\Bigg(\frac{\nu_{\text{1}}}{\nu_{\text{2}}}\Bigg)^{-\alpha_{\text{s}}\gamma}\Bigg(\frac{\mathcal{F_{\nu_{\text{1}}\text{,min}}}}{\mathcal{F_{\nu_{\text{2}}\text{,min}}}}\Bigg)^{\gamma+\text{1}},
\label{Equation:DetectionRate}
\end{equation}
where $\gamma$\,$=$\,--2.3\,$\pm$\,0.4 is a power law index that describes FRB~180916.J0158+65's differential energy distribution~\citep{CHIME+2020b}, and $\mathcal{F_{\nu_{\text{1}}\text{,min}}}$ and $\mathcal{F_{\nu_{\text{2}}\text{,min}}}$ are the fluence detection thresholds at frequencies $\nu_{\text{1}}$ and $\nu_{\text{2}}$, respectively.

In Table~\ref{Table:Table5}, we list the predicted burst rates above the average 6$\sigma$~fluence detection thresholds during our observations in each of these time intervals using Equation~\eqref{Equation:DetectionRate}. The total exposure times at $S$-band and $X$-band with DSS-63 during subintervals 1--3 were $t_{\text{1}}$\,$=$\,11.9\,hr, $t_{\text{2}}$\,$=$\,15.7\,hr, and $t_{\text{3}}$\,$=$\,8.5\,hr, respectively. The $L$-band observations with DSS-14 spanned a total of 11.0\,hr, which occurred entirely during subinterval~1. Therefore, the expected number of detectable bursts in each of these frequency bands and time intervals is given by $N$\,$=$\,$r$\,$\times$\,$t$, where $r$ is the burst rate and $t$ is the total exposure time in each time interval. These values are also provided in Table~\ref{Table:Table5}. The uncertainties on the burst rates and number of expected bursts are dominated by the large errors on the statistical spectral index, $\alpha_{\text{s}}$, which is poorly constrained for FRB~180916.J0158+65.

These estimates are all consistent with our observations, which suggest that fewer or fainter bursts are detectable from FRB~180916.J0158+65 at higher radio frequencies ($\gtrsim$\,2\,GHz). Intrinsic and extrinsic effects may also affect the observed burst rates. For example, FRB~180916.J0158+65 exhibits periods of apparent inactivity during some activity cycles and, like many other repeating~FRB sources, emits narrowband radio bursts (e.g., see~\citealt{CHIME+2020b}). This behavior is often highly variable between observing frequencies.

We also consider the following less likely explanations for the lack of radio bursts observed from FRB~180916.J0158+65 during our radio monitoring campaign: (1)~On average, FRB~180916.J0158+65 emits bursts that have narrower widths at higher frequencies (e.g., see~\citealt{Chawla+2020, CHIME+2020b, Marcote+2020}). While it is conceivable that bursts with narrow pulse widths ($\sim$10\,$\mu$s) and high peak luminosities may not have been detected in the data recorded at $S$-band and $X$-band with the pulsar backend, if their fluences did not exceed the fluence detection thresholds listed in Table~\ref{Table:Table2}, no evidence of such emission was found during a preliminary inspection of simultaneously recorded high time resolution baseband data during times when bursts were detected between 400--800\,MHz with CHIME/FRB. Additionally, our search for bursts using data from the pulsar backend was sufficiently sensitive to detect bursts with similar characteristics as most of those detected at 1.7\,GHz with the EVN~\citep{Marcote+2020}. (2)~If FRB~180916.J0158+65 has a smaller duty cycle at high frequencies ($\gtrsim$\,2\,GHz), with a burst rate that is comparable to the rate reported at lower frequencies (e.g., see~\citealt{CHIME+2020b}), then it is possible that many bursts would not be detected unless the observations were performed during activity phases when the source was producing high frequency radio emission. However, we detected no radio bursts during our observations, which spanned $\sim$40\% of the FRB~180916.J0158+65's 16.35\,d activity cycle and $\sim$50\% of the $\pm$2.7\,d interval around the peak of the source's activity phase (see Figure~\ref{Figure:Figure2}), including 8~times when bursts were simultaneously detected from the source in the 400--800\,MHz frequency band (see Table~\ref{Table:Table4} and Figure~\ref{Figure:Figure3}).

\subsection{Progenitor Models}
\label{Section:ProgenitorModels}

Several progenitor models have been proposed that may explain the emission characteristics of repeating~FRBs and the periodicity observed thus far from a subset of these sources. For example, the apparent 16.35\,d period observed from FRB~180916.J0158+65 could be attributed to eccentric orbital motion of a binary system comprised of a neutron star and a massive O/B-type companion~\citep{CHIME+2020b}, where the emitter could be either a radio pulsar or magnetar. Other models suggest that the apparent periodicity may arise from free-free absorption in the wind of the massive companion, which could lead to modulated emission that is dependent on the orbital phase~\citep{Lyutikov+2020}. The optical depth in the companion's homogeneous, isothermal wind is expected to decrease with increasing frequency. Under these assumptions, the activity window is predicted to be longer at higher radio frequencies. While some radio bursts detected from FRB~180916.J0158+65 at higher frequencies have been seen to occur earlier in the activity window~\citep{CHIME+2020b, Aggarwal+2020a}, compared to those detected at lower frequencies, the source's activity had not been well-characterized above $\sim$2\,GHz across a wide range of activity phases, until now. The cadence of our observations of FRB~180916.J0158+65 and the absence of radio emission at 2.3 and 8.4\,GHz during times when the source was known to be active together suggest that the duration of the activity window is either not strongly frequency-dependent or narrower at high radio frequencies and possibly systematically shifted in phase.

The binary comb model, proposed by~\citet{Ioka+2020}, also associates the 16.35\,d period observed from FRB~180916.J0158+65 with the orbital period of an interacting neutron star binary system. In this scenario, the FRB emission is produced by a highly magnetized pulsar, whose emission is funneled by the strong wind of a millisecond pulsar or massive stellar companion. The observed activity window then coincides with times when the funnel is directed toward Earth.

Alternatively, FRB~180916.J0158+65's periodicity has been interpreted as arising from the precession of a flaring magnetar or magnetized neutron star~\citep{Levin+2020, Tong+2020, Yang+2020, Zanazzi+2020}. \citet{Levin+2020} showed that a hyperactive magnetar, driven by a high rate of ambipolar diffusion in the core, can have a precession period ranging from weeks to months, and the viscous damping timescale is orders of magnitude longer than the magnetar's age. Other models suggest that the 16.35\,d period observed from FRB~180916.J0158+65 may originate from a 1--10\,kyr old magnetar with an ultra-long rotation period and a high internal magnetic field ($B_{\text{int}}$\,$\gtrsim$\,10$^{\text{16}}$\,G) at birth~\citep{Beniamini+2020} or the precession of a jet produced by the accretion disk of a massive black hole~\citep{Katz2020}.

Recently, tentative evidence of a $\sim$157\,d period, with a duty cycle of 56\%, was discovered during a search for periodicity in the arrival times of radio bursts from FRB~121102~\citep{Rajwade+2020b}. This long-term periodicity was also observed by~\citet{Aggarwal+2020b} and~\citet{Cruces+2020}, and they both found a period that was consistent with the measurement reported in \citet{Rajwade+2020b}. Both of the $S$-band bursts shown in Figure~\ref{Figure:Figure1} occurred at an activity phase of $\sim$0.53, determined by folding the burst arrival times modulo the $\sim$157\,d period and using the reference time, $t_{\text{ref}}$\,$=$\,MJD~58200, defined in~\citet{Rajwade+2020b}. These bursts were both detected near the peak of the activity window, defined at activity phase~0.5. Although it is not yet clear whether the $\sim$157\,d period is astrophysical in origin, the radio bursts presented here from FRB~121102 are consistent with the observed periodicity. If the $\sim$157\,d period is determined to be astrophysical, it suggests that other repeating~FRB sources may display underlying periodic behavior, which would provide important clues about the nature of their progenitors.

Although FRBs and Galactic radio magnetars share some similar emission characteristics, such as prominent frequency-structure in their radio pulses (e.g., see~\citealt{Gajjar+2018, Pearlman+2018, Hessels+2019, Maan+2019, Pearlman+2020}) and extended periods of radio inactivity (e.g., see~\citealt{Scholz+2017b, Camilo+2018, Levin+2019}), there are also notable differences. High frequency radio emission is a hallmark feature of radio magnetars. Bright radio pulses have been detected from several radio magnetars in the Milky Way at record high frequencies, well above 100\,GHz in some cases~(e.g., see~\citealt{Camilo+2007, Torne+2017, Torne+2020}), but no FRB~source has yet been detected at radio frequencies above 8\,GHz. However, to date, very few searches for FRBs have been carried out at such high radio frequencies~\citep{Gajjar+2018, Majid+2020}.

Radio bursts from FRB sources are also considerably more energetic, on average, than radio pulses from Galactic magnetars~\citep{Pearlman+2018, Hessels+2019, Pearlman+2020}. The recent detection of a bright radio pulse from SGR~1935+2154 has demonstrated that magnetars can produce radio bursts with a wide range of luminosities, spanning from luminosities comparable to radio pulsars up to about a factor of $\sim$10 lower than the luminosities of bursts observed from FRB~180916.J0158+65~\citep{Bochenek+2020, CHIME+2020a, Kirsten+2020, Zhang+2020}. However, the apparent burst repetition rate of FRB-like bursts from Galactic magnetars is inconsistent with the repetition rates observed from extragalactic repeating FRBs~\citep{Lu+2020a, Margalit+2020}. For example, for at least the past 7~years, FRB~121102 has been producing radio bursts that are $\sim$10$^{\text{10}}$ times more energetic than the radio pulses observed from Galactic radio magnetars~\citep{Spitler+2014, Spitler+2016}. No magnetar in the Milky Way has been found to exhibit a similar level of activity. If repeating FRBs are produced by active magnetars, this suggests that they must differ in some way from the Galactic magnetar population.

The short temporal durations of FRB pulses indicate that they originate from compact emitting regions (e.g., see~\citealt{Michilli+2018a, Cho+2020}). However, the peculiar spectral properties between sources and the downward-drifting subpulse behavior frequently observed from repeating~FRBs are not yet well-explained. Wideband, multifrequency observations of FRBs are therefore crucial for understanding their spectral properties and the underlying emission mechanism(s) powering these sources.


\section{Summary and Conclusions}
\label{Section:Conclusions}

We carried out long-term monitoring observations of two repeating~FRB sources, FRB~121102 and FRB~180916.J0158+65, at high radio frequencies using the DSN's 70-m radio telescopes (DSS-63 and DSS-14). Radio observations of FRB~121102 were performed between 2019~September~19 (MJD~58745) and 2020~February~11 (MJD~58890), and we observed FRB~180916.J0158+65 between 2019~September~19 (MJD~58745) and 2020~May~14 (MJD~58983). Most of our observations were performed simultaneously at $S$-band (2.3\,GHz) and $X$-band (8.4\,GHz) with DSS-63. We also observed FRB~180916.J0158+65 at $L$-band (1.5\,GHz) during 8~separate epochs using DSS-14.

We detected 2~radio bursts from FRB~121102 at $S$-band (see~Figure~\ref{Figure:Figure1}) during periods when the source was predicted to be active. No accompanying radio emission was observed at $X$-band during the same time. Both of these bursts occurred near the near the peak of the activity window predicted by~\citet{Rajwade+2020b}. These detections provide additional examples of the source's narrowband emission behavior at high radio frequencies, which cannot be explained by Galactic DISS~\citep{Majid+2020}.

Our radio observations of FRB~180916.J0158+65 span a wide range of activity phases (see Figure~\ref{Figure:Figure2}), including several cycles when the source was known to be active at lower frequencies. Until now, FRB~180916.J0158+65's emission behavior had not been explored at frequencies above $\sim$2\,GHz across a wide range of activity phases. During our observing campaign, 8~radio bursts were detected from FRB~180916.J0158+65 in the 400--800\,MHz band by CHIME/FRB~\citep{CHIME+2020b}\textsuperscript{\ref{FootnoteCHIMEWebsite}}. The arrival times of these bursts occurred during times when we were simultaneously observing the source with DSS-14 at $L$-band or DSS-63 at $S$/$X$-band. However, no evidence of radio emission was found in our observing bands during these times (e.g., see Figure~\ref{Figure:Figure3}), and no radio bursts were detected from FRB~180916.J0158+65 during any other times. These observations further demonstrate that FRB~180916.J0158+65's emission process is strongly frequency-dependent~(e.g., see~\citealt{Chawla+2020, CHIME+2020b}), with fewer or fainter bursts emitted from the source at high frequencies. We also find that the radio bursts detected from FRB~180916.J0158+65 are band-limited and have less spectral occupancy at high frequencies. Therefore, the apparent activity of the source, like FRB~121102, strongly depends on the choice of radio observing frequency.


\section{Acknowledgments}

We thank Professor Vicky Kaspi and the CHIME/FRB collaboration for their support of these observations and for providing the CHIME/FRB data used in Figure~\ref{Figure:Figure3}. We also thank the reviewer for valuable comments and suggestions.

A.B.P. acknowledges support by the Department of Defense~(DoD) through the National Defense Science and Engineering Graduate~(NDSEG) Fellowship Program and by the National Science Foundation~(NSF) Graduate Research Fellowship under Grant~No.~\text{DGE-1144469}. J.W.T.H. acknowledges funding from an NWO Vici fellowship.

We thank the Jet Propulsion Laboratory's Spontaneous Concept Research and Technology Development program for supporting this work. We also thank Dr.~Stephen~Lichten for providing programmatic support. In addition, we are grateful to the DSN scheduling team (Hernan~Diaz, George~Martinez, and Carleen~Ward) and the GDSCC and MDSCC operations staff for scheduling and carrying out these observations.

A portion of this research was performed at the Jet Propulsion Laboratory, California Institute of Technology and the Caltech campus, under a Research and Technology Development Grant through a contract with the National Aeronautics and Space Administration. U.S. government sponsorship is acknowledged.


\section*{ORCID iDs}
\label{Section:OrcidIDs}

\noindent
Aaron~B.~Pearlman~\href{https:/orcid.org/0000-0002-8912-0732}{https:/orcid.org/0000-0002-8912-0732} \\
\noindent
Walid~A.~Majid~\href{https:/orcid.org/0000-0002-4694-4221}{https:/orcid.org/0000-0002-4694-4221} \\
\noindent
Thomas~A.~Prince~\href{https:/orcid.org/0000-0002-8850-3627}{https:/orcid.org/0000-0002-8850-3627} \\
\noindent
Kenzie~Nimmo~\href{https://orcid.org/0000-0003-0510-0740}{https://orcid.org/0000-0003-0510-0740} \\
\noindent
Jason~W.~T.~Hessels~\href{https://orcid.org/0000-0003-2317-1446}{https://orcid.org/0000-0003-2317-1446} \\
\noindent
Charles~J.~Naudet~\href{https://orcid.org/0000-0001-6898-0533}{https://orcid.org/0000-0001-6898-0533}
\noindent
Jonathon~Kocz~\href{https:/orcid.org/0000-0003-0249-7586}{https:/orcid.org/0000-0003-0249-7586}


\bibliography{references}

\begin{thebibliography}{}
\providecommand\natexlab[1]{#1}
\providecommand\JournalTitle[1]{#1}

\bibitem[{{Agarwal} {et~al.}(2020){Agarwal}, {Aggarwal}, {Burke-Spolaor},
  {Lorimer}, \& {Garver-Daniels}}]{Agarwal+2020}
{Agarwal}, D., {Aggarwal}, K., {Burke-Spolaor}, S., {Lorimer}, D.~R., \&
  {Garver-Daniels}, N. 2020,
  \href{http://dx.doi.org/10.1093/mnras/staa1856}{\JournalTitle{\mnras}, 497,
  1661}

\bibitem[{{Aggarwal} {et~al.}(2020){Aggarwal}, {Law}, {Burke-Spolaor}, {Bower},
  {Butler}, {Demorest}, {Linford}, \& {Lazio}}]{Aggarwal+2020b}
{Aggarwal}, K., {Law}, C.~J., {Burke-Spolaor}, S., {et~al.} 2020,
  \href{http://dx.doi.org/10.3847/2515-5172/ab9f33}{\JournalTitle{Research
  Notes of the American Astronomical Society}, 4, 94}

\bibitem[{{Aggarwal} \& {Realfast Collaboration}(2020)}]{Aggarwal+2020a}
{Aggarwal}, K. \& {Realfast Collaboration}. 2020, \JournalTitle{The
  Astronomer's Telegram}, 13664, 1

\bibitem[{{Bannister} {et~al.}(2019){Bannister}, {Deller}, {Phillips},
  {Macquart}, {Prochaska}, {Tejos}, {Ryder}, {Sadler}, {Shannon}, {Simha},
  {Day}, {McQuinn}, {North-Hickey}, {Bhandari}, {Arcus}, {Bennert}, {Burchett},
  {Bouwhuis}, {Dodson}, {Ekers}, {Farah}, {Flynn}, {James}, {Kerr}, {Lenc},
  {Mahony}, {O'Meara}, {Os{\l}owski}, {Qiu}, {Treu}, {U}, {Bateman}, {Bock},
  {Bolton}, {Brown}, {Bunton}, {Chippendale}, {Cooray}, {Cornwell}, {Gupta},
  {Hayman}, {Kesteven}, {Koribalski}, {MacLeod}, {McClure-Griffiths},
  {Neuhold}, {Norris}, {Pilawa}, {Qiao}, {Reynolds}, {Roxby}, {Shimwell},
  {Voronkov}, \& {Wilson}}]{Bannister+2019}
{Bannister}, K.~W., {Deller}, A.~T., {Phillips}, C., {et~al.} 2019,
  \href{http://dx.doi.org/10.1126/science.aaw5903}{\JournalTitle{Science}, 365,
  565}

\bibitem[{{Beloborodov}(2017)}]{Beloborodov2017}
{Beloborodov}, A.~M. 2017,
  \href{http://dx.doi.org/10.3847/2041-8213/aa78f3}{\JournalTitle{\apjl}, 843,
  L26}

\bibitem[{{Beniamini} {et~al.}(2020){Beniamini}, {Wadiasingh}, \&
  {Metzger}}]{Beniamini+2020}
{Beniamini}, P., {Wadiasingh}, Z., \& {Metzger}, B.~D. 2020,
  \href{http://dx.doi.org/10.1093/mnras/staa1783}{\JournalTitle{\mnras}, 496,
  3390}

\bibitem[{{Bhandari} {et~al.}(2020){Bhandari}, {Sadler}, {Prochaska}, {Simha},
  {Ryder}, {Marnoch}, {Bannister}, {Macquart}, {Flynn}, {Shannon}, {Tejos},
  {Corro-Guerra}, {Day}, {Deller}, {Ekers}, {Lopez}, {Mahony}, {Nu{\~n}ez}, \&
  {Phillips}}]{Bhandari+2020}
{Bhandari}, S., {Sadler}, E.~M., {Prochaska}, J.~X., {et~al.} 2020,
  \href{http://dx.doi.org/10.3847/2041-8213/ab672e}{\JournalTitle{\apjl}, 895,
  L37}

\bibitem[{{Bochenek} {et~al.}(2020){Bochenek}, {Ravi}, {Belov}, {Hallinan},
  {Kocz}, {Kulkarni}, \& {McKenna}}]{Bochenek+2020}
{Bochenek}, C.~D., {Ravi}, V., {Belov}, K.~V., {et~al.} 2020,
  \href{http://dx.doi.org/10.1038/s41586-020-2872-x}{\JournalTitle{\nat}, 587,
  59}

\bibitem[{{Camilo} {et~al.}(2007){Camilo}, {Ransom}, {Pe{\~n}alver},
  {Karastergiou}, {van Kerkwijk}, {Durant}, {Halpern}, {Reynolds}, {Thum},
  {Helfand }, {Zimmerman}, \& {Cognard}}]{Camilo+2007}
{Camilo}, F., {Ransom}, S.~M., {Pe{\~n}alver}, J., {et~al.} 2007,
  \href{http://dx.doi.org/10.1086/521548}{\JournalTitle{\apj}, 669, 561}

\bibitem[{{Camilo} {et~al.}(2018){Camilo}, {Scholz}, {Serylak}, {Buchner},
  {Merryfield}, {Kaspi}, {Archibald}, {Bailes}, {Jameson}, {van Straten},
  {Sarkissian}, {Reynolds}, {Johnston}, {Hobbs}, {Abbott}, {Adam}, {Adams},
  {Alberts}, {Andreas}, {Asad}, {Baker}, {Baloyi}, {Bauermeister}, {Baxana},
  {Bennett}, {Bernardi}, {Booisen}, {Booth}, {Botha}, {Boyana}, {Brederode},
  {Burger}, {Cheetham}, {Conradie}, {Conradie}, {Davidson}, {De Bruin}, {de
  Swardt}, {de Villiers}, {de Villiers}, {de Villiers}, {de Villiers}, {De
  Waal}, {Dikgale}, {du Toit}, {du Toit}, {Esterhuyse}, {Fanaroff}, {Fataar},
  {Foley}, {Foster}, {Fourie}, {Gamatham}, {Gatsi}, {Geschke}, {Goedhart},
  {Grobler}, {Gumede}, {Hlakola}, {Hokwana}, {Hoorn}, {Horn}, {Horrell},
  {Hugo}, {Isaacson}, {Jacobs}, {Jansen van Rensburg}, {Jonas}, {Jordaan},
  {Joubert}, {Joubert}, {J{\'o}zsa}, {Julie}, {Julius}, {Kapp}, {Karastergiou},
  {Karels}, {Kariseb}, {Karuppusamy}, {Kasper}, {Knox-Davies}, {Koch},
  {Kotz{\'e}}, {Krebs}, {Kriek}, {Kriel}, {Kusel}, {Lamoor}, {Lehmensiek},
  {Liebenberg}, {Liebenberg}, {Lord}, {Lunsky}, {Mabombo}, {Macdonald},
  {Macfarlane}, {Madisa}, {Mafhungo}, {Magnus}, {Magozore}, {Mahgoub}, {Main},
  {Makhathini}, {Malan}, {Malgas}, {Manley}, {Manzini}, {Marais}, {Marais},
  {Marais}, {Maree}, {Martens}, {Matshawule}, {Matthysen}, {Mauch}, {McNally},
  {Merry}, {Millenaar}, {Mjikelo}, {Mkhabela}, {Mnyand u}, {Moeng}, {Mokone},
  {Monama}, {Montshiwa}, {Moss}, {Mphego}, {New}, {Ngcebetsha}, {Ngoasheng},
  {Niehaus}, {Ntuli}, {Nzama}, {Obies}, {Obrocka}, {Ockards}, {Olyn}, {Oozeer},
  {Otto}, {Padayachee}, {Passmoor}, {Patel}, {Paula}, {Peens-Hough},
  {Pholoholo}, {Prozesky}, {Rakoma}, {Ramaila}, {Rammala}, {Ramudzuli},
  {Rasivhaga}, {Ratcliffe}, {Reader}, {Renil}, {Richter}, {Robyntjies},
  {Rosekrans}, {Rust}, {Salie}, {Sambu}, {Schollar}, {Schwardt}, {Seranyane},
  {Sethosa}, {Sharpe}, {Siebrits}, {Sirothia}, {Slabber}, {Smirnov}, {Smith},
  {Sofeya}, {Songqumase}, {Spann}, {Stappers}, {Steyn}, {Steyn}, {Strong},
  {Struthers}, {Stuart}, {Sunnylall}, {Swart}, {Taljaard}, {Tasse}, {Taylor},
  {Theron}, {Thondikulam}, {Thorat}, {Tiplady}, {Toruvanda}, {van Aardt}, {van
  Balla}, {van den Heever}, {van der Byl}, {van der Merwe}, {van der Merwe},
  {van Niekerk}, {van Rooyen}, {van Staden}, {van Tonder}, {van Wyk}, {Wait},
  {Walker}, {Wallace}, {Welz}, {Williams}, {Xaia}, {Young}, \&
  {Zitha}}]{Camilo+2018}
{Camilo}, F., {Scholz}, P., {Serylak}, M., {et~al.} 2018,
  \href{http://dx.doi.org/10.3847/1538-4357/aab35a}{\JournalTitle{\apj}, 856,
  180}

\bibitem[{{Chatterjee} {et~al.}(2017){Chatterjee}, {Law}, {Wharton},
  {Burke-Spolaor}, {Hessels}, {Bower}, {Cordes}, {Tendulkar}, {Bassa},
  {Demorest}, {Butler}, {Seymour}, {Scholz}, {Abruzzo}, {Bogdanov}, {Kaspi},
  {Keimpema}, {Lazio}, {Marcote}, {McLaughlin}, {Paragi}, {Ransom}, {Rupen},
  {Spitler}, \& {van Langevelde}}]{Chatterjee+2017}
{Chatterjee}, S., {Law}, C.~J., {Wharton}, R.~S., {et~al.} 2017,
  \href{http://dx.doi.org/10.1038/nature20797}{\JournalTitle{\nat}, 541, 58}

\bibitem[{{Chawla} {et~al.}(2020){Chawla}, {Andersen}, {Bhardwaj}, {Fonseca},
  {Josephy}, {Kaspi}, {Michilli}, {Pleunis}, {Bandura}, {Bassa}, {Boyle},
  {Brar}, {Cassanelli}, {Cubranic}, {Dobbs}, {Dong}, {Gaensler}, {Good},
  {Hessels}, {Land ecker}, {Leung}, {Li}, {Lin}, {Masui}, {Mckinven},
  {Mena-Parra}, {Merryfield}, {Meyers}, {Naidu}, {Ng}, {Patel},
  {Rafiei-Ravandi}, {Rahman}, {Sanghavi}, {Scholz}, {Shin}, {Smith}, {Stairs},
  {Tendulkar}, \& {Vanderlinde}}]{Chawla+2020}
{Chawla}, P., {Andersen}, B.~C., {Bhardwaj}, M., {et~al.} 2020,
  \href{http://dx.doi.org/10.3847/2041-8213/ab96bf}{\JournalTitle{\apjl}, 896,
  L41}

\bibitem[{{Cho} {et~al.}(2020){Cho}, {Macquart}, {Shannon}, {Deller},
  {Morrison}, {Ekers}, {Bannister}, {Farah}, {Qiu}, {Sammons}, {Bailes},
  {Bhandari}, {Day}, {James}, {Phillips}, {Prochaska}, \& {Tuthill}}]{Cho+2020}
{Cho}, H., {Macquart}, J.-P., {Shannon}, R.~M., {et~al.} 2020,
  \href{http://dx.doi.org/10.3847/2041-8213/ab7824}{\JournalTitle{\apjl}, 891,
  L38}

\bibitem[{{Cordes} \& {Chatterjee}(2019)}]{Cordes+2019}
{Cordes}, J.~M. \& {Chatterjee}, S. 2019,
  \href{http://dx.doi.org/10.1146/annurev-astro-091918-104501}{\JournalTitle{\araa},
  57, 417}

\bibitem[{{Cordes} \& {Lazio}(2002)}]{Cordes+2002}
{Cordes}, J.~M. \& {Lazio}, T.~J.~W. 2002, \JournalTitle{arXiv e-prints}, astro

\bibitem[{{Cordes} \& {Rickett}(1998)}]{Cordes+1998}
{Cordes}, J.~M. \& {Rickett}, B.~J. 1998,
  \href{http://dx.doi.org/10.1086/306358}{\JournalTitle{\apj}, 507, 846}

\bibitem[{{Cordes} {et~al.}(2017){Cordes}, {Wasserman}, {Hessels}, {Lazio},
  {Chatterjee}, \& {Wharton}}]{Cordes+2017}
{Cordes}, J.~M., {Wasserman}, I., {Hessels}, J.~W.~T., {et~al.} 2017,
  \href{http://dx.doi.org/10.3847/1538-4357/aa74da}{\JournalTitle{\apj}, 842,
  35}

\bibitem[{{Cruces} {et~al.}(2020){Cruces}, {Spitler}, {Scholz}, {Lynch},
  {Seymour}, {Hessels}, {Gouiff{\'e}s}, {Hilmarsson}, {Kramer}, \&
  {Munjal}}]{Cruces+2020}
{Cruces}, M., {Spitler}, L.~G., {Scholz}, P., {et~al.} 2020,
  \href{http://dx.doi.org/10.1093/mnras/staa3223}{\JournalTitle{\mnras}},
  \href{http://arxiv.org/abs/2008.03461}{{\sffamily arXiv:2008.03461}}

\bibitem[{{Day} {et~al.}(2020){Day}, {Deller}, {Shannon}, {Qiu}, {Bannister},
  {Bhandari}, {Ekers}, {Flynn}, {James}, {Macquart}, {Mahony}, {Phillips}, \&
  {Xavier Prochaska}}]{Day+2020}
{Day}, C.~K., {Deller}, A.~T., {Shannon}, R.~M., {et~al.} 2020,
  \href{http://dx.doi.org/10.1093/mnras/staa2138}{\JournalTitle{\mnras}, 497,
  3335}

\bibitem[{{Fonseca} {et~al.}(2020){Fonseca}, {Andersen}, {Bhardwaj}, {Chawla},
  {Good}, {Josephy}, {Kaspi}, {Masui}, {Mckinven}, {Michilli}, {Pleunis},
  {Shin}, {Tendulkar}, {Bandura}, {Boyle}, {Brar}, {Cassanelli}, {Cubranic},
  {Dobbs}, {Dong}, {Gaensler}, {Hinshaw}, {Land ecker}, {Leung}, {Li}, {Lin},
  {Mena-Parra}, {Merryfield}, {Naidu}, {Ng}, {Patel}, {Pen}, {Rafiei-Ravandi},
  {Rahman}, {Ransom}, {Scholz}, {Smith}, {Stairs}, {Vanderlinde}, {Yadav}, \&
  {Zwaniga}}]{Fonseca+2020}
{Fonseca}, E., {Andersen}, B.~C., {Bhardwaj}, M., {et~al.} 2020,
  \href{http://dx.doi.org/10.3847/2041-8213/ab7208}{\JournalTitle{\apjl}, 891,
  L6}

\bibitem[{{Gajjar} {et~al.}(2018){Gajjar}, {Siemion}, {Price}, {Law},
  {Michilli}, {Hessels}, {Chatterjee}, {Archibald}, {Bower}, {Brinkman},
  {Burke-Spolaor}, {Cordes}, {Croft}, {Enriquez}, {Foster}, {Gizani},
  {Hellbourg}, {Isaacson}, {Kaspi}, {Lazio}, {Lebofsky}, {Lynch}, {MacMahon},
  {McLaughlin}, {Ransom}, {Scholz}, {Seymour}, {Spitler}, {Tendulkar},
  {Werthimer}, \& {Zhang}}]{Gajjar+2018}
{Gajjar}, V., {Siemion}, A.~P.~V., {Price}, D.~C., {et~al.} 2018,
  \href{http://dx.doi.org/10.3847/1538-4357/aad005}{\JournalTitle{\apj}, 863,
  2}

\bibitem[{{Gourdji} {et~al.}(2019){Gourdji}, {Michilli}, {Spitler}, {Hessels},
  {Seymour}, {Cordes}, \& {Chatterjee}}]{Gourdji+2019}
{Gourdji}, K., {Michilli}, D., {Spitler}, L.~G., {et~al.} 2019,
  \href{http://dx.doi.org/10.3847/2041-8213/ab1f8a}{\JournalTitle{\apjl}, 877,
  L19}

\bibitem[{{Heintz} {et~al.}(2020){Heintz}, {Prochaska}, {Simha}, {Platts},
  {Fong}, {Tejos}, {Ryder}, {Aggarwal}, {Bhandari}, {Day}, {Deller},
  {Kilpatrick}, {Law}, {Macquart}, {Mannings}, {Marnoch}, {Sadler}, \&
  {Shannon}}]{Heintz+2020}
{Heintz}, K.~E., {Prochaska}, J.~X., {Simha}, S., {et~al.} 2020,
  \JournalTitle{arXiv e-prints}, arXiv:2009.10747

\bibitem[{{Hessels} {et~al.}(2019){Hessels}, {Spitler}, {Seymour}, {Cordes},
  {Michilli}, {Lynch}, {Gourdji}, {Archibald}, {Bassa}, {Bower}, {Chatterjee},
  {Connor}, {Crawford}, {Deneva}, {Gajjar}, {Kaspi}, {Keimpema}, {Law},
  {Marcote}, {McLaughlin}, {Paragi}, {Petroff}, {Ransom}, {Scholz}, {Stappers},
  \& {Tendulkar}}]{Hessels+2019}
{Hessels}, J.~W.~T., {Spitler}, L.~G., {Seymour}, A.~D., {et~al.} 2019,
  \href{http://dx.doi.org/10.3847/2041-8213/ab13ae}{\JournalTitle{\apjl}, 876,
  L23}

\bibitem[{{Hotan} {et~al.}(2004){Hotan}, {van Straten}, \&
  {Manchester}}]{Hotan+2004}
{Hotan}, A.~W., {van Straten}, W., \& {Manchester}, R.~N. 2004,
  \href{http://dx.doi.org/10.1071/AS04022}{\JournalTitle{\pasa}, 21, 302}

\bibitem[{{Houben} {et~al.}(2019){Houben}, {Spitler}, {ter Veen}, {Rachen},
  {Falcke}, \& {Kramer}}]{Houben+2019}
{Houben}, L.~J.~M., {Spitler}, L.~G., {ter Veen}, S., {et~al.} 2019,
  \href{http://dx.doi.org/10.1051/0004-6361/201833875}{\JournalTitle{\aap},
  623, A42}

\bibitem[{{Ioka} \& {Zhang}(2020)}]{Ioka+2020}
{Ioka}, K. \& {Zhang}, B. 2020,
  \href{http://dx.doi.org/10.3847/2041-8213/ab83fb}{\JournalTitle{\apjl}, 893,
  L26}

\bibitem[{{Josephy} {et~al.}(2019){Josephy}, {Chawla}, {Fonseca}, {Ng},
  {Patel}, {Pleunis}, {Scholz}, {Andersen}, {Bandura}, {Bhardwaj}, {Boyce},
  {Boyle}, {Brar}, {Cubranic}, {Dobbs}, {Gaensler}, {Gill}, {Giri}, {Good},
  {Halpern}, {Hinshaw}, {Kaspi}, {Landecker}, {Lang}, {Lin}, {Masui},
  {Mckinven}, {Mena-Parra}, {Merryfield}, {Michilli}, {Milutinovic}, {Naidu},
  {Pen}, {Rafiei-Ravand i}, {Rahman}, {Ransom}, {Renard}, {Siegel}, {Smith},
  {Stairs}, {Tendulkar}, {Vanderlinde}, {Yadav}, \& {Zwaniga}}]{Josephy+2019}
{Josephy}, A., {Chawla}, P., {Fonseca}, E., {et~al.} 2019,
  \href{http://dx.doi.org/10.3847/2041-8213/ab2c00}{\JournalTitle{\apjl}, 882,
  L18}

\bibitem[{{Katz}(2020)}]{Katz2020}
{Katz}, J.~I. 2020,
  \href{http://dx.doi.org/10.1093/mnrasl/slaa038}{\JournalTitle{\mnras}, 494,
  L64}

\bibitem[{{Kirsten} {et~al.}(2020){Kirsten}, {Snelders}, {Jenkins}, {Nimmo},
  {van den Eijnden}, {Hessels}, {Gawronski}, \& {Yang}}]{Kirsten+2020}
{Kirsten}, F., {Snelders}, M., {Jenkins}, M., {et~al.} 2020,
  \JournalTitle{arXiv e-prints}, arXiv:2007.05101

\bibitem[{{Kumar} {et~al.}(2019){Kumar}, {Shannon}, {Os{\l}owski}, {Qiu},
  {Bhandari}, {Farah}, {Flynn}, {Kerr}, {Lorimer}, {Macquart}, {Ng},
  {Phillips}, {Price}, \& {Spiewak}}]{Kumar+2019}
{Kumar}, P., {Shannon}, R.~M., {Os{\l}owski}, S., {et~al.} 2019,
  \href{http://dx.doi.org/10.3847/2041-8213/ab5b08}{\JournalTitle{\apjl}, 887,
  L30}

\bibitem[{{Kumar} {et~al.}(2020){Kumar}, {Shannon}, {Flynn}, {Os{\l}owski},
  {Bhandari}, {Day}, {Deller}, {Farah}, {Kaczmarek}, {Kerr}, {Phillips},
  {Price}, {Qiu}, \& {Thyagarajan}}]{Kumar+2020}
{Kumar}, P., {Shannon}, R.~M., {Flynn}, C., {et~al.} 2020, \JournalTitle{arXiv
  e-prints}, arXiv:2009.01214

\bibitem[{{Law} {et~al.}(2017){Law}, {Abruzzo}, {Bassa}, {Bower},
  {Burke-Spolaor}, {Butler}, {Cantwell}, {Carey}, {Chatterjee}, {Cordes},
  {Demorest}, {Dowell}, {Fender}, {Gourdji}, {Grainge}, {Hessels}, {Hickish},
  {Kaspi}, {Lazio}, {McLaughlin}, {Michilli}, {Mooley}, {Perrott}, {Ransom},
  {Razavi-Ghods}, {Rupen}, {Scaife}, {Scott}, {Scholz}, {Seymour}, {Spitler},
  {Stovall}, {Tendulkar}, {Titterington}, {Wharton}, \& {Williams}}]{Law+2017}
{Law}, C.~J., {Abruzzo}, M.~W., {Bassa}, C.~G., {et~al.} 2017,
  \href{http://dx.doi.org/10.3847/1538-4357/aa9700}{\JournalTitle{\apj}, 850,
  76}

\bibitem[{{Law} {et~al.}(2020){Law}, {Butler}, {Prochaska}, {Zackay},
  {Burke-Spolaor}, {Mannings}, {Tejos}, {Josephy}, {Andersen}, {Chawla},
  {Heintz}, {Aggarwal}, {Bower}, {Demorest}, {Kilpatrick}, {Lazio}, {Linford},
  {Mckinven}, {Tendulkar}, \& {Simha}}]{Law+2020}
{Law}, C.~J., {Butler}, B.~J., {Prochaska}, J.~X., {et~al.} 2020,
  \href{http://dx.doi.org/10.3847/1538-4357/aba4ac}{\JournalTitle{\apj}, 899,
  161}

\bibitem[{{Levin} {et~al.}(2019){Levin}, {Lyne}, {Desvignes}, {Eatough},
  {Karuppusamy}, {Kramer}, {Mickaliger}, {Stappers}, \&
  {Weltevrede}}]{Levin+2019}
{Levin}, L., {Lyne}, A.~G., {Desvignes}, G., {et~al.} 2019,
  \href{http://dx.doi.org/10.1093/mnras/stz2074}{\JournalTitle{\mnras}, 488,
  5251}

\bibitem[{{Levin} {et~al.}(2020){Levin}, {Beloborodov}, \&
  {Bransgrove}}]{Levin+2020}
{Levin}, Y., {Beloborodov}, A.~M., \& {Bransgrove}, A. 2020,
  \href{http://dx.doi.org/10.3847/2041-8213/ab8c4c}{\JournalTitle{\apjl}, 895,
  L30}

\bibitem[{{Lu} \& {Kumar}(2018)}]{Lu+2018}
{Lu}, W. \& {Kumar}, P. 2018,
  \href{http://dx.doi.org/10.1093/mnras/sty716}{\JournalTitle{\mnras}, 477,
  2470}

\bibitem[{{Lu} {et~al.}(2020{\natexlab{a}}){Lu}, {Kumar}, \&
  {Zhang}}]{Lu+2020a}
{Lu}, W., {Kumar}, P., \& {Zhang}, B. 2020{\natexlab{a}},
  \href{http://dx.doi.org/10.1093/mnras/staa2450}{\JournalTitle{\mnras}, 498,
  1397}

\bibitem[{{Lu} {et~al.}(2020{\natexlab{b}}){Lu}, {Piro}, \&
  {Waxman}}]{Lu+2020b}
{Lu}, W., {Piro}, A.~L., \& {Waxman}, E. 2020{\natexlab{b}},
  \href{http://dx.doi.org/10.1093/mnras/staa2397}{\JournalTitle{\mnras}},
  \href{http://arxiv.org/abs/2003.12581}{{\sffamily arXiv:2003.12581}}

\bibitem[{{Luo} {et~al.}(2020){Luo}, {Wang}, {Men}, {Zhang}, {Jiang}, {Xu},
  {Wang}, {Lee}, {Han}, {Zhang}, {Caballero}, {Chen}, {Chen}, {Gan}, {Guo},
  {Hao}, {Huang}, {Jiang}, {Li}, {Li}, {Li}, {Luo}, {Pan}, {Pei}, {Qian},
  {Sun}, {Wang}, {Wang}, {Wen}, {Xu}, {Xu}, {Yan}, {Yan}, {Yu}, {Yuan},
  {Zhang}, \& {Zhu}}]{Luo+2020}
{Luo}, R., {Wang}, B.~J., {Men}, Y.~P., {et~al.} 2020,
  \href{http://dx.doi.org/10.1038/s41586-020-2827-2}{\JournalTitle{\nat}, 586,
  693}

\bibitem[{{Lyutikov} {et~al.}(2020){Lyutikov}, {Barkov}, \&
  {Giannios}}]{Lyutikov+2020}
{Lyutikov}, M., {Barkov}, M.~V., \& {Giannios}, D. 2020,
  \href{http://dx.doi.org/10.3847/2041-8213/ab87a4}{\JournalTitle{\apjl}, 893,
  L39}

\bibitem[{{Maan} {et~al.}(2019){Maan}, {Joshi}, {Surnis}, {Bagchi}, \&
  {Manoharan}}]{Maan+2019}
{Maan}, Y., {Joshi}, B.~C., {Surnis}, M.~P., {Bagchi}, M., \& {Manoharan},
  P.~K. 2019,
  \href{http://dx.doi.org/10.3847/2041-8213/ab3a47}{\JournalTitle{\apjl}, 882,
  L9}

\bibitem[{{Macquart} {et~al.}(2020){Macquart}, {Prochaska}, {McQuinn},
  {Bannister}, {Bhandari}, {Day}, {Deller}, {Ekers}, {James}, {Marnoch},
  {Os{\l}owski}, {Phillips}, {Ryder}, {Scott}, {Shannon}, \&
  {Tejos}}]{Macquart+2020}
{Macquart}, J.~P., {Prochaska}, J.~X., {McQuinn}, M., {et~al.} 2020,
  \href{http://dx.doi.org/10.1038/s41586-020-2300-2}{\JournalTitle{\nat}, 581,
  391}

\bibitem[{{Majid} {et~al.}(2017){Majid}, {Pearlman}, {Dobreva}, {Horiuchi},
  {Kocz}, {Lippuner}, \& {Prince}}]{Majid+2017}
{Majid}, W.~A., {Pearlman}, A.~B., {Dobreva}, T., {et~al.} 2017,
  \href{http://dx.doi.org/10.3847/2041-8213/834/1/L2}{\JournalTitle{\apjl},
  834, L2}

\bibitem[{{Majid} {et~al.}(2020){Majid}, {Pearlman}, {Nimmo}, {Hessels},
  {Prince}, {Naudet}, {Kocz}, \& {Horiuchi}}]{Majid+2020}
{Majid}, W.~A., {Pearlman}, A.~B., {Nimmo}, K., {et~al.} 2020,
  \href{http://dx.doi.org/10.3847/2041-8213/ab9a4a}{\JournalTitle{\apjl}, 897,
  L4}

\bibitem[{{Marcote} {et~al.}(2017){Marcote}, {Paragi}, {Hessels}, {Keimpema},
  {van Langevelde}, {Huang}, {Bassa}, {Bogdanov}, {Bower}, {Burke-Spolaor},
  {Butler}, {Campbell}, {Chatterjee}, {Cordes}, {Demorest}, {Garrett}, {Ghosh},
  {Kaspi}, {Law}, {Lazio}, {McLaughlin}, {Ransom}, {Salter}, {Scholz},
  {Seymour}, {Siemion}, {Spitler}, {Tendulkar}, \& {Wharton}}]{Marcote+2017}
{Marcote}, B., {Paragi}, Z., {Hessels}, J.~W.~T., {et~al.} 2017,
  \href{http://dx.doi.org/10.3847/2041-8213/834/2/L8}{\JournalTitle{\apjl},
  834, L8}

\bibitem[{{Marcote} {et~al.}(2020){Marcote}, {Nimmo}, {Hessels}, {Tendulkar},
  {Bassa}, {Paragi}, {Keimpema}, {Bhardwaj}, {Karuppusamy}, {Kaspi}, {Law},
  {Michilli}, {Aggarwal}, {Andersen}, {Archibald}, {Bandura}, {Bower}, {Boyle},
  {Brar}, {Burke-Spolaor}, {Butler}, {Cassanelli}, {Chawla}, {Demorest},
  {Dobbs}, {Fonseca}, {Giri}, {Good}, {Gourdji}, {Josephy}, {Kirichenko},
  {Kirsten}, {Landecker}, {Lang}, {Lazio}, {Li}, {Lin}, {Linford}, {Masui},
  {Mena-Parra}, {Naidu}, {Ng}, {Patel}, {Pen}, {Pleunis}, {Rafiei-Ravandi},
  {Rahman}, {Renard}, {Scholz}, {Siegel}, {Smith}, {Stairs}, {Vanderlinde}, \&
  {Zwaniga}}]{Marcote+2020}
{Marcote}, B., {Nimmo}, K., {Hessels}, J.~W.~T., {et~al.} 2020,
  \href{http://dx.doi.org/10.1038/s41586-019-1866-z}{\JournalTitle{\nat}, 577,
  190}

\bibitem[{{Margalit} {et~al.}(2020){Margalit}, {Beniamini}, {Sridhar}, \&
  {Metzger}}]{Margalit+2020}
{Margalit}, B., {Beniamini}, P., {Sridhar}, N., \& {Metzger}, B.~D. 2020,
  \href{http://dx.doi.org/10.3847/2041-8213/abac57}{\JournalTitle{\apjl}, 899,
  L27}

\bibitem[{{Marthi} {et~al.}(2020){Marthi}, {Gautam}, {Li}, {Lin}, {Main},
  {Naidu}, {Pen}, \& {Wharton}}]{Marthi+2020}
{Marthi}, V.~R., {Gautam}, T., {Li}, D.~Z., {et~al.} 2020,
  \href{http://dx.doi.org/10.1093/mnrasl/slaa148}{\JournalTitle{\mnras}, 499,
  L16}

\bibitem[{{McLaughlin} \& {Cordes}(2003)}]{McLaughlin+2003}
{McLaughlin}, M.~A. \& {Cordes}, J.~M. 2003,
  \href{http://dx.doi.org/10.1086/378232}{\JournalTitle{\apj}, 596, 982}

\bibitem[{{Metzger} {et~al.}(2017){Metzger}, {Berger}, \&
  {Margalit}}]{Metzger+2017}
{Metzger}, B.~D., {Berger}, E., \& {Margalit}, B. 2017,
  \href{http://dx.doi.org/10.3847/1538-4357/aa633d}{\JournalTitle{\apj}, 841,
  14}

\bibitem[{{Michilli} {et~al.}(2018){Michilli}, {Seymour}, {Hessels}, {Spitler},
  {Gajjar}, {Archibald}, {Bower}, {Chatterjee}, {Cordes}, {Gourdji}, {Heald},
  {Kaspi}, {Law}, {Sobey}, {Adams}, {Bassa}, {Bogdanov}, {Brinkman},
  {Demorest}, {Fernand ez}, {Hellbourg}, {Lazio}, {Lynch}, {Maddox}, {Marcote},
  {McLaughlin}, {Paragi}, {Ransom}, {Scholz}, {Siemion}, {Tendulkar}, {van
  Rooy}, {Wharton}, \& {Whitlow}}]{Michilli+2018a}
{Michilli}, D., {Seymour}, A., {Hessels}, J.~W.~T., {et~al.} 2018,
  \href{http://dx.doi.org/10.1038/nature25149}{\JournalTitle{\nat}, 553, 182}

\bibitem[{{Nimmo} {et~al.}(2020){Nimmo}, {Hessels}, {Keimpema}, {Archibald},
  {Cordes}, {Karuppusamy}, {Kirsten}, {Li}, {Marcote}, \&
  {Paragi}}]{Nimmo+2020}
{Nimmo}, K., {Hessels}, J.~W.~T., {Keimpema}, A., {et~al.} 2020,
  \JournalTitle{arXiv e-prints}, arXiv:2010.05800

\bibitem[{{Pearlman} {et~al.}(2019{\natexlab{a}}){Pearlman}, {Majid}, \&
  {Prince}}]{Pearlman+2019a}
{Pearlman}, A.~B., {Majid}, W.~A., \& {Prince}, T.~A. 2019{\natexlab{a}},
  \href{http://dx.doi.org/10.1155/2019/6325183}{\JournalTitle{Advances in
  Astronomy}, 2019, 6325183}

\bibitem[{{Pearlman} {et~al.}(2018){Pearlman}, {Majid}, {Prince}, {Kocz}, \&
  {Horiuchi}}]{Pearlman+2018}
{Pearlman}, A.~B., {Majid}, W.~A., {Prince}, T.~A., {Kocz}, J., \& {Horiuchi},
  S. 2018,
  \href{http://dx.doi.org/10.3847/1538-4357/aade4d}{\JournalTitle{\apj}, 866,
  160}

\bibitem[{{Pearlman} {et~al.}(2019{\natexlab{b}}){Pearlman}, {Majid}, {Prince},
  {Naudet}, {Hessels}, {Nimmo}, {Kocz}, \& {Horiuchi}}]{Pearlman+2019b}
{Pearlman}, A.~B., {Majid}, W.~A., {Prince}, T.~A., {et~al.}
  2019{\natexlab{b}}, \JournalTitle{The Astronomer's Telegram}, 13235, 1

\bibitem[{{Pearlman} {et~al.}(2020){Pearlman}, {Majid}, {Prince}, {Ray},
  {Kocz}, {Horiuchi}, {Naudet}, {G{\"u}ver}, {Enoto}, {Arzoumanian},
  {Gendreau}, \& {Ho}}]{Pearlman+2020}
{Pearlman}, A.~B., {Majid}, W.~A., {Prince}, T.~A., {et~al.} 2020,
  \JournalTitle{arXiv e-prints}, arXiv:2005.08410

\bibitem[{{Petroff} {et~al.}(2019){Petroff}, {Hessels}, \&
  {Lorimer}}]{Petroff+2019}
{Petroff}, E., {Hessels}, J.~W.~T., \& {Lorimer}, D.~R. 2019,
  \href{http://dx.doi.org/10.1007/s00159-019-0116-6}{\JournalTitle{\aapr}, 27,
  4}

\bibitem[{{Petroff} {et~al.}(2016){Petroff}, {Barr}, {Jameson}, {Keane},
  {Bailes}, {Kramer}, {Morello}, {Tabbara}, \& {van Straten}}]{Petroff+2016}
{Petroff}, E., {Barr}, E.~D., {Jameson}, A., {et~al.} 2016,
  \href{http://dx.doi.org/10.1017/pasa.2016.35}{\JournalTitle{\pasa}, 33, e045}

\bibitem[{{Pilia} {et~al.}(2020){Pilia}, {Burgay}, {Possenti}, {Ridolfi},
  {Gajjar}, {Corongiu}, {Perrodin}, {Bernardi}, {Naldi}, {Pupillo},
  {Ambrosino}, {Bianchi}, {Burtovoi}, {Casella}, {Casentini}, {Cecconi},
  {Ferrigno}, {Fiori}, {Gendreau}, {Ghedina}, {Naletto}, {Nicastro}, {Ochner},
  {Palazzi}, {Panessa}, {Papitto}, {Pittori}, {Rea}, {Castillo}, {Savchenko},
  {Setti}, {Tavani}, {Trois}, {Trudu}, {Turatto}, {Ursi}, {Verrecchia}, \&
  {Zampieri}}]{Pilia+2020}
{Pilia}, M., {Burgay}, M., {Possenti}, A., {et~al.} 2020,
  \href{http://dx.doi.org/10.3847/2041-8213/ab96c0}{\JournalTitle{\apjl}, 896,
  L40}

\bibitem[{{Platts} {et~al.}(2019){Platts}, {Weltman}, {Walters}, {Tendulkar},
  {Gordin}, \& {Kandhai}}]{Platts+2019}
{Platts}, E., {Weltman}, A., {Walters}, A., {et~al.} 2019,
  \href{http://dx.doi.org/10.1016/j.physrep.2019.06.003}{\JournalTitle{\physrep},
  821, 1}

\bibitem[{{Prochaska} {et~al.}(2019){Prochaska}, {Macquart}, {McQuinn},
  {Simha}, {Shannon}, {Day}, {Marnoch}, {Ryder}, {Deller}, {Bannister},
  {Bhandari}, {Bordoloi}, {Bunton}, {Cho}, {Flynn}, {Mahony}, {Phillips},
  {Qiu}, \& {Tejos}}]{Prochaska+2019}
{Prochaska}, J.~X., {Macquart}, J.-P., {McQuinn}, M., {et~al.} 2019,
  \href{http://dx.doi.org/10.1126/science.aay0073}{\JournalTitle{Science}, 366,
  231}

\bibitem[{{Rajwade} {et~al.}(2020{\natexlab{a}}){Rajwade}, {Mickaliger},
  {Stappers}, {Bassa}, {Breton}, {Karastergiou}, \& {Keane}}]{Rajwade+2020a}
{Rajwade}, K.~M., {Mickaliger}, M.~B., {Stappers}, B.~W., {et~al.}
  2020{\natexlab{a}},
  \href{http://dx.doi.org/10.1093/mnras/staa616}{\JournalTitle{\mnras}, 493,
  4418}

\bibitem[{{Rajwade} {et~al.}(2020{\natexlab{b}}){Rajwade}, {Mickaliger},
  {Stappers}, {Morello}, {Agarwal}, {Bassa}, {Breton}, {Caleb}, {Karastergiou},
  {Keane}, \& {Lorimer}}]{Rajwade+2020b}
{Rajwade}, K.~M., {Mickaliger}, M.~B., {Stappers}, B.~W., {et~al.}
  2020{\natexlab{b}},
  \href{http://dx.doi.org/10.1093/mnras/staa1237}{\JournalTitle{\mnras}, 495,
  3551}

\bibitem[{{Ransom}(2001)}]{Ransom+2001}
{Ransom}, S.~M. 2001, PhD thesis, Harvard University

\bibitem[{{Ravi}(2019)}]{Ravi2019}
{Ravi}, V. 2019,
  \href{http://dx.doi.org/10.1038/s41550-019-0831-y}{\JournalTitle{Nature
  Astronomy}, 3, 928}

\bibitem[{{Ravi} \& {Loeb}(2019)}]{Ravi+2019b}
{Ravi}, V. \& {Loeb}, A. 2019,
  \href{http://dx.doi.org/10.3847/1538-4357/ab0748}{\JournalTitle{\apj}, 874,
  72}

\bibitem[{{Ravi} {et~al.}(2019){Ravi}, {Catha}, {D'Addario}, {Djorgovski},
  {Hallinan}, {Hobbs}, {Kocz}, {Kulkarni}, {Shi}, {Vedantham}, {Weinreb}, \&
  {Woody}}]{Ravi+2019a}
{Ravi}, V., {Catha}, M., {D'Addario}, L., {et~al.} 2019,
  \href{http://dx.doi.org/10.1038/s41586-019-1389-7}{\JournalTitle{\nat}, 572,
  352}

\bibitem[{{Sand} {et~al.}(2020){Sand}, {Gajjar}, {Pilia}, {Kudale}, {Joshi},
  {Jagtap}, {Ray}, {Deshpande}, {Bijay}, {Dey}, {Kalita}, {Bandyopadhyay},
  {Jena}, {Bhattacharya}, {Waratkar}, {Wagle}, {Singha}, {Bagchi}, {Surnis},
  {Bhat}, {Mishra}, {Konar}, \& {Maan}}]{Sand+2020}
{Sand}, K.~R., {Gajjar}, V., {Pilia}, M., {et~al.} 2020, \JournalTitle{The
  Astronomer's Telegram}, 13781, 1

\bibitem[{{Scholz} {et~al.}(2016){Scholz}, {Spitler}, {Hessels}, {Chatterjee},
  {Cordes}, {Kaspi}, {Wharton}, {Bassa}, {Bogdanov}, {Camilo}, {Crawford},
  {Deneva}, {van Leeuwen}, {Lynch}, {Madsen}, {McLaughlin}, {Mickaliger},
  {Parent}, {Patel}, {Ransom}, {Seymour}, {Stairs}, {Stappers}, \&
  {Tendulkar}}]{Scholz+2016}
{Scholz}, P., {Spitler}, L.~G., {Hessels}, J.~W.~T., {et~al.} 2016,
  \href{http://dx.doi.org/10.3847/1538-4357/833/2/177}{\JournalTitle{\apj},
  833, 177}

\bibitem[{{Scholz} {et~al.}(2017{\natexlab{a}}){Scholz}, {Bogdanov}, {Hessels},
  {Lynch}, {Spitler}, {Bassa}, {Bower}, {Burke-Spolaor}, {Butler},
  {Chatterjee}, {Cordes}, {Gourdji}, {Kaspi}, {Law}, {Marcote}, {McLaughlin},
  {Michilli}, {Paragi}, {Ransom}, {Seymour}, {Tendulkar}, \&
  {Wharton}}]{Scholz+2017a}
{Scholz}, P., {Bogdanov}, S., {Hessels}, J.~W.~T., {et~al.} 2017{\natexlab{a}},
  \href{http://dx.doi.org/10.3847/1538-4357/aa8456}{\JournalTitle{\apj}, 846,
  80}

\bibitem[{{Scholz} {et~al.}(2017{\natexlab{b}}){Scholz}, {Camilo},
  {Sarkissian}, {Reynolds}, {Levin}, {Bailes}, {Burgay}, {Johnston}, {Kramer},
  \& {Possenti}}]{Scholz+2017b}
{Scholz}, P., {Camilo}, F., {Sarkissian}, J., {et~al.} 2017{\natexlab{b}},
  \href{http://dx.doi.org/10.3847/1538-4357/aa73de}{\JournalTitle{\apj}, 841,
  126}

\bibitem[{{Scholz} {et~al.}(2020){Scholz}, {Cook}, {Cruces}, {Hessels},
  {Kaspi}, {Majid}, {Naidu}, {Pearlman}, {Spitler}, {Bandura}, {Bhardwaj},
  {Cassanelli}, {Chawla}, {Gaensler}, {Good}, {Josephy}, {Karuppusamy},
  {Keimpema}, {Kirichenko}, {Kirsten}, {Kocz}, {Leung}, {Marcote}, {Masui},
  {Mena-Parra}, {Merryfield}, {Michilli}, {Naudet}, {Nimmo}, {Pleunis},
  {Prince}, {Rafiei-Ravandi}, {Rahman}, {Shin}, {Smith}, {Stairs}, {Tendulkar},
  \& {Vanderlinde}}]{Scholz+2020}
{Scholz}, P., {Cook}, A., {Cruces}, M., {et~al.} 2020,
  \href{http://dx.doi.org/10.3847/1538-4357/abb1a8}{\JournalTitle{\apj}, 901,
  165}

\bibitem[{{Shannon} {et~al.}(2018){Shannon}, {Macquart}, {Bannister}, {Ekers},
  {James}, {Os{\l}owski}, {Qiu}, {Sammons}, {Hotan}, {Voronkov}, {Beresford},
  {Brothers}, {Brown}, {Bunton}, {Chippendale}, {Haskins}, {Leach},
  {Marquarding}, {McConnell}, {Pilawa}, {Sadler}, {Troup}, {Tuthill},
  {Whiting}, {Allison}, {Anderson}, {Bell}, {Collier}, {G{\"u}rkan}, {Heald},
  \& {Riseley}}]{Shannon+2018}
{Shannon}, R.~M., {Macquart}, J.~P., {Bannister}, K.~W., {et~al.} 2018,
  \href{http://dx.doi.org/10.1038/s41586-018-0588-y}{\JournalTitle{\nat}, 562,
  386}

\bibitem[{{Spitler} {et~al.}(2014){Spitler}, {Cordes}, {Hessels}, {Lorimer},
  {McLaughlin}, {Chatterjee}, {Crawford}, {Deneva}, {Kaspi}, {Wharton},
  {Allen}, {Bogdanov}, {Brazier}, {Camilo}, {Freire}, {Jenet},
  {Karako-Argaman}, {Knispel}, {Lazarus}, {Lee}, {van Leeuwen}, {Lynch},
  {Ransom}, {Scholz}, {Siemens}, {Stairs}, {Stovall}, {Swiggum},
  {Venkataraman}, {Zhu}, {Aulbert}, \& {Fehrmann}}]{Spitler+2014}
{Spitler}, L.~G., {Cordes}, J.~M., {Hessels}, J.~W.~T., {et~al.} 2014,
  \href{http://dx.doi.org/10.1088/0004-637X/790/2/101}{\JournalTitle{\apj},
  790, 101}

\bibitem[{{Spitler} {et~al.}(2016){Spitler}, {Scholz}, {Hessels}, {Bogdanov},
  {Brazier}, {Camilo}, {Chatterjee}, {Cordes}, {Crawford}, {Deneva}, {Ferdman},
  {Freire}, {Kaspi}, {Lazarus}, {Lynch}, {Madsen}, {McLaughlin}, {Patel},
  {Ransom}, {Seymour}, {Stairs}, {Stappers}, {van Leeuwen}, \&
  {Zhu}}]{Spitler+2016}
{Spitler}, L.~G., {Scholz}, P., {Hessels}, J.~W.~T., {et~al.} 2016,
  \href{http://dx.doi.org/10.1038/nature17168}{\JournalTitle{\nat}, 531, 202}

\bibitem[{{Spitler} {et~al.}(2018){Spitler}, {Herrmann}, {Bower}, {Chatterjee},
  {Cordes}, {Hessels}, {Kramer}, {Michilli}, {Scholz}, {Seymour}, \&
  {Siemion}}]{Spitler+2018}
{Spitler}, L.~G., {Herrmann}, W., {Bower}, G.~C., {et~al.} 2018,
  \href{http://dx.doi.org/10.3847/1538-4357/aad332}{\JournalTitle{\apj}, 863,
  150}

\bibitem[{{Tendulkar} {et~al.}(2017){Tendulkar}, {Bassa}, {Cordes}, {Bower},
  {Law}, {Chatterjee}, {Adams}, {Bogdanov}, {Burke-Spolaor}, {Butler},
  {Demorest}, {Hessels}, {Kaspi}, {Lazio}, {Maddox}, {Marcote}, {McLaughlin},
  {Paragi}, {Ransom}, {Scholz}, {Seymour}, {Spitler}, {van Langevelde}, \&
  {Wharton}}]{Tendulkar+2017}
{Tendulkar}, S.~P., {Bassa}, C.~G., {Cordes}, J.~M., {et~al.} 2017,
  \href{http://dx.doi.org/10.3847/2041-8213/834/2/L7}{\JournalTitle{\apjl},
  834, L7}

\bibitem[{{Tendulkar} {et~al.}(2020){Tendulkar}, {Gil de Paz}, {Kirichenko},
  {Hessels}, {Bhardwaj}, {{\'A}vila}, {Bassa}, {Chawla}, {Fonseca}, {Kaspi},
  {Keimpema}, {Kirsten}, {Lazio}, {Marcote}, {Masui}, {Nimmo}, {Paragi},
  {Rahman}, {Reverte Pay{\'a}}, {Scholz}, \& {Stairs}}]{Tendulkar+2020}
{Tendulkar}, S.~P., {Gil de Paz}, A., {Kirichenko}, A.~Y., {et~al.} 2020,
  \JournalTitle{arXiv e-prints}, arXiv:2011.03257

\bibitem[{{The CHIME/FRB Collaboration} {et~al.}(2018){The CHIME/FRB
  Collaboration}, {Amiri}, {Bandura}, {Berger}, {Bhardwaj}, {Boyce}, {Boyle},
  {Brar}, {Burhanpurkar}, {Chawla}, {Chowdhury}, {Cliche}, {Cranmer},
  {Cubranic}, {Deng}, {Denman}, {Dobbs}, {Fandino}, {Fonseca}, {Gaensler},
  {Giri}, {Gilbert}, {Good}, {Guliani}, {Halpern}, {Hinshaw}, {H{\"o}fer},
  {Josephy}, {Kaspi}, {Landecker}, {Lang}, {Liao}, {Masui}, {Mena-Parra},
  {Naidu}, {Newburgh}, {Ng}, {Patel}, {Pen}, {Pinsonneault-Marotte}, {Pleunis},
  {Rafiei Ravandi}, {Ransom}, {Renard}, {Scholz}, {Sigurdson}, {Siegel},
  {Smith}, {Stairs}, {Tendulkar}, {Vand erlinde}, \& {Wiebe}}]{CHIME+2018}
{The CHIME/FRB Collaboration}, {Amiri}, M., {Bandura}, K., {et~al.} 2018,
  \href{http://dx.doi.org/10.3847/1538-4357/aad188}{\JournalTitle{\apj}, 863,
  48}

\bibitem[{{The CHIME/FRB Collaboration} {et~al.}(2019{\natexlab{a}}){The
  CHIME/FRB Collaboration}, {Amiri}, {Bandura}, {Bhardwaj}, {Boubel}, {Boyce},
  {Boyle}, {. Brar}, {Burhanpurkar}, {Cassanelli}, {Chawla}, {Cliche},
  {Cubranic}, {Deng}, {Denman}, {Dobbs}, {Fandino}, {Fonseca}, {Gaensler},
  {Gilbert}, {Gill}, {Giri}, {Good}, {Halpern}, {Hanna}, {Hill}, {Hinshaw},
  {H{\"o}fer}, {Josephy}, {Kaspi}, {Landecker}, {Lang}, {Lin}, {Masui},
  {Mckinven}, {Mena-Parra}, {Merryfield}, {Michilli}, {Milutinovic}, {Moatti},
  {Naidu}, {Newburgh}, {Ng}, {Patel}, {Pen}, {Pinsonneault-Marotte}, {Pleunis},
  {Rafiei-Ravandi}, {Rahman}, {Ransom}, {Renard}, {Scholz}, {Shaw}, {Siegel},
  {Smith}, {Stairs}, {Tendulkar}, {Tretyakov}, {Vanderlinde}, \&
  {Yadav}}]{CHIME+2019a}
{The CHIME/FRB Collaboration}, {Amiri}, M., {Bandura}, K., {et~al.}
  2019{\natexlab{a}},
  \href{http://dx.doi.org/10.1038/s41586-018-0864-x}{\JournalTitle{\nat}, 566,
  235}

\bibitem[{{The CHIME/FRB Collaboration} {et~al.}(2019{\natexlab{b}}){The
  CHIME/FRB Collaboration}, {Andersen}, {Bandura}, {Bhardwaj}, {Boubel},
  {Boyce}, {Boyle}, {Brar}, {Cassanelli}, {Chawla}, {Cubranic}, {Deng},
  {Dobbs}, {Fandino}, {Fonseca}, {Gaensler}, {Gilbert}, {Giri}, {Good},
  {Halpern}, {Hill}, {Hinshaw}, {H{\"o}fer}, {Josephy}, {Kaspi}, {Kothes},
  {Landecker}, {Lang}, {Li}, {Lin}, {Masui}, {Mena-Parra}, {Merryfield},
  {Mckinven}, {Michilli}, {Milutinovic}, {Naidu}, {Newburgh}, {Ng}, {Patel},
  {Pen}, {Pinsonneault-Marotte}, {Pleunis}, {Rafiei-Ravandi}, {Rahman},
  {Ransom}, {Renard}, {Scholz}, {Siegel}, {Singh}, {Smith}, {Stairs},
  {Tendulkar}, {Tretyakov}, {Vanderlinde}, {Yadav}, \& {Zwaniga}}]{CHIME+2019b}
{The CHIME/FRB Collaboration}, {Andersen}, B.~C., {Bandura}, K., {et~al.}
  2019{\natexlab{b}},
  \href{http://dx.doi.org/10.3847/2041-8213/ab4a80}{\JournalTitle{\apjl}, 885,
  L24}

\bibitem[{{The CHIME/FRB Collaboration} {et~al.}(2020{\natexlab{a}}){The
  CHIME/FRB Collaboration}, {Andersen}, {Bandura}, {Bhardwaj}, {Bij}, {Boyce},
  {Boyle}, {Brar}, {Cassanelli}, {Chawla}, {Chen}, {Cliche}, {Cook},
  {Cubranic}, {Curtin}, {Denman}, {Dobbs}, {Dong}, {Fandino}, {Fonseca},
  {Gaensler}, {Giri}, {Good}, {Halpern}, {Hill}, {Hinshaw}, {H{"o}fer},
  {Josephy}, {Kania}, {Kaspi}, {Landecker}, {Leung}, {Li}, {Lin}, {Masui},
  {McKinven}, {Mena-Parra}, {Merryfield}, {Meyers}, {Michilli}, {Milutinovic},
  {Mirhosseini}, {M{"u}nchmeyer}, {Naidu}, {Newburgh}, {Ng}, {Patel}, {Pen},
  {Pinsonneault-Marotte}, {Pleunis}, {Quine}, {Rafiei-Ravandi}, {Rahman},
  {Ransom}, {Renard}, {Sanghavi}, {Scholz}, {Shaw}, {Shin}, {Siegel}, {Singh},
  {Smegal}, {Smith}, {Stairs}, {Tan}, {Tendulkar}, {Tretyakov}, {Vanderlinde},
  {Wang}, {Wulf}, \& {Zwaniga}}]{CHIME+2020a}
{The CHIME/FRB Collaboration}, {Andersen}, B.~C., {Bandura}, K.~M., {et~al.}
  2020{\natexlab{a}},
  \href{http://dx.doi.org/10.1038/s41586-020-2863-y}{\JournalTitle{\nat}, 587,
  54}

\bibitem[{{The CHIME/FRB Collaboration} {et~al.}(2020{\natexlab{b}}){The
  CHIME/FRB Collaboration}, {Amiri}, {Andersen}, {Bandura}, {Bhardwaj},
  {Boyle}, {Brar}, {Chawla}, {Chen}, {Cliche}, {Cubranic}, {Deng}, {Denman},
  {Dobbs}, {Dong}, {Fand ino}, {Fonseca}, {Gaensler}, {Giri}, {Good},
  {Halpern}, {Hessels}, {Hill}, {H{\"o}fer}, {Josephy}, {Kania}, {Karuppusamy},
  {Kaspi}, {Keimpema}, {Kirsten}, {Landecker}, {Lang}, {Leung}, {Li}, {Lin},
  {Marcote}, {Masui}, {Mckinven}, {Mena-Parra}, {Merryfield}, {Michilli},
  {Milutinovic}, {Mirhosseini}, {Naidu}, {Newburgh}, {Ng}, {Nimmo}, {Paragi},
  {Patel}, {Pen}, {Pinsonneault-Marotte}, {Pleunis}, {Rafiei-Ravandi},
  {Rahman}, {Ransom}, {Renard}, {Sanghavi}, {Scholz}, {Shaw}, {Shin}, {Siegel},
  {Singh}, {Smegal}, {Smith}, {Stairs}, {Tendulkar}, {Tretyakov},
  {Vanderlinde}, {Wang}, {Wang}, {Wulf}, {Yadav}, \& {Zwaniga}}]{CHIME+2020b}
{The CHIME/FRB Collaboration}, {Amiri}, M., {Andersen}, B.~C., {et~al.}
  2020{\natexlab{b}},
  \href{http://dx.doi.org/10.1038/s41586-020-2398-2}{\JournalTitle{\nat}, 582,
  351}

\bibitem[{{Tong} {et~al.}(2020){Tong}, {Wang}, \& {Wang}}]{Tong+2020}
{Tong}, H., {Wang}, W., \& {Wang}, H.-G. 2020,
  \href{http://dx.doi.org/10.1088/1674-4527/20/9/142}{\JournalTitle{Research in
  Astronomy and Astrophysics}, 20, 142}

\bibitem[{{Torne} {et~al.}(2017){Torne}, {Desvignes}, {Eatough}, {Karuppusamy},
  {Paubert}, {Kramer}, {Cognard}, {Champion}, \& {Spitler}}]{Torne+2017}
{Torne}, P., {Desvignes}, G., {Eatough}, R.~P., {et~al.} 2017,
  \href{http://dx.doi.org/10.1093/mnras/stw2757}{\JournalTitle{\mnras}, 465,
  242}

\bibitem[{{Torne} {et~al.}(2020){Torne}, {Mac{\'\i}as-P{\'e}rez}, {Ladjelate},
  {Ritacco}, {S{\'a}nchez-Portal}, {Berta}, {Paubert}, {Calvo}, {Desvignes},
  {Karuppusamy}, {Navarro}, {John}, {S{\'a}nchez}, {Pe{\~n}alver}, {Kramer}, \&
  {Schuster}}]{Torne+2020}
{Torne}, P., {Mac{\'\i}as-P{\'e}rez}, J., {Ladjelate}, B., {et~al.} 2020,
  \href{http://dx.doi.org/10.1051/0004-6361/202038504}{\JournalTitle{\aap},
  640, L2}

\bibitem[{{Yang} \& {Zou}(2020)}]{Yang+2020}
{Yang}, H. \& {Zou}, Y.-C. 2020,
  \href{http://dx.doi.org/10.3847/2041-8213/ab800f}{\JournalTitle{\apjl}, 893,
  L31}

\bibitem[{{Zanazzi} \& {Lai}(2020)}]{Zanazzi+2020}
{Zanazzi}, J.~J. \& {Lai}, D. 2020,
  \href{http://dx.doi.org/10.3847/2041-8213/ab7cdd}{\JournalTitle{\apjl}, 892,
  L15}

\bibitem[{{Zhang} {et~al.}(2020){Zhang}, {Jiang}, {Men}, {Wang}, {Xu}, {Xu},
  {Niu}, {Zhou}, {Guan}, {Han}, {Jiang}, {Lee}, {Li}, {Lin}, {Niu}, {Wang},
  {Wang}, {Xu}, {Yu}, {Zhang}, \& {Zhu}}]{Zhang+2020}
{Zhang}, C.~F., {Jiang}, J.~C., {Men}, Y.~P., {et~al.} 2020, \JournalTitle{The
  Astronomer's Telegram}, 13699, 1

\bibitem[{{Zhang} {et~al.}(2018){Zhang}, {Gajjar}, {Foster}, {Siemion},
  {Cordes}, {Law}, \& {Wang}}]{Zhang+2018}
{Zhang}, Y.~G., {Gajjar}, V., {Foster}, G., {et~al.} 2018,
  \href{http://dx.doi.org/10.3847/1538-4357/aadf31}{\JournalTitle{\apj}, 866,
  149}

\end{thebibliography}


\newpage

\begin{deluxetable*}{c|ccccccc}
	
	\tablenum{1}
	\tabletypesize{\small}
	\tablecolumns{8}
	\tablewidth{0pt}
	\tablecaption{\textsc{Multiwavelength Radio Observations of FRB~121102}}
	\tablehead{
		Telescope$^{\mathrm{a}}$ & 
		Start Time$^{\mathrm{b}}$ & 
		Exposure &
		Center & 
		Bandwidth$^{\mathrm{c}}$ &
		Number of &
		Time &
		6$\sigma$ Fluence \\
		& 
		&
		Time & 
		Frequency & 
		&
		Polarizations$^{\mathrm{d}}$ &
		Resolution & 
		Threshold ($\mathcal{F_\text{min}}$)$^{\mathrm{e}}$ \\
		&
		(UTC) &
		(s) &
		(GHz) & 
		(MHz) & 
		& 
		(ms) & 
		$\big(\text{Jy\,ms}\,\sqrt{(w/1\,\text{ms})}\big)$
	}
	\startdata
	\multirow{40}{*}{DSS-63} & \multirow{2}{*}{2019-09-19 00:55:25} & \multirow{2}{*}{4379.964} & 2.25 & 120.14 & 1 & \multirow{2}{*}{2.2075} & 0.34 \\
	&  &  & 8.38 & 395.68 & 2 &  & 0.14 \\
	\cline{2-8} \xrowht{5pt}
	& \multirow{2}{*}{2019-09-24 00:07:07} & \multirow{2}{*}{5339.983} & 2.25 & 120.14 & 1 & \multirow{2}{*}{2.2075} & 0.35 \\
	&  &  & 8.38 & 395.68 & 2 &  & 0.15 \\
	\cline{2-8} \xrowht{5pt}
	& \multirow{2}{*}{\textbf{2019-09-28 00:02:05}} & \multirow{2}{*}{5159.923} & 2.25 & 120.14 & 1 & \multirow{2}{*}{2.2075} & 0.35 \\
	&  &  & 8.38 & 395.68 & 2 &  & 0.14 \\
	\cline{2-8} \xrowht{5pt}
	& \multirow{2}{*}{\textbf{2019-09-28 23:57:05}} & \multirow{2}{*}{5339.983} & 2.25 & 120.14 & 1 & \multirow{2}{*}{2.2075} & 0.35 \\
	&  &  & 8.38 & 395.68 & 2 &  & 0.14 \\
	\cline{2-8} \xrowht{5pt}
	& \multirow{2}{*}{2019-10-19 20:52:15} & \multirow{2}{*}{3155.671} & 2.25 & 120.14 & 1 & \multirow{2}{*}{0.2759} & 0.39 \\
	&  &  & 8.38 & 395.68 & 2 &  & 0.17 \\
	\cline{2-8} \xrowht{5pt}
	& \multirow{2}{*}{2019-10-21 22:17:07} & \multirow{2}{*}{5197.230} & 2.25 & 120.14 & 1 & \multirow{2}{*}{0.2759} & 0.35 \\
	&  &  & 8.38 & 395.68 & 2 &  & 0.14 \\
	\cline{2-8} \xrowht{5pt}
	& \multirow{2}{*}{2019-11-09 21:12:15} & \multirow{2}{*}{3596.984} & 2.25 & 110.40 & \multirow{2}{*}{2} & \multirow{2}{*}{1.0240} & 0.27 \\
	&  &  & 8.38 & 395.68 &  &  & 0.15 \\
	\cline{2-8} \xrowht{5pt}
	& \multirow{2}{*}{2019-11-20 21:56:53} & \multirow{2}{*}{5459.935} & 2.25 & 110.40 & \multirow{2}{*}{2} & \multirow{2}{*}{2.2075} & 0.25 \\
	&  &  & 8.38 & 395.68 &  &  & 0.13 \\
	\cline{2-8} \xrowht{5pt}
	& \multirow{2}{*}{2019-11-21 03:14:05} & \multirow{2}{*}{2799.930} & 2.25 & 110.40 & \multirow{2}{*}{2} & \multirow{2}{*}{2.2075} & 0.24 \\
	&  &  & 8.38 & 395.68 &  &  & 0.13 \\
	\cline{2-8} \xrowht{5pt}
	& \multirow{2}{*}{2019-11-25 20:00:45} & \multirow{2}{*}{10641.985} & 2.25 & 110.40 & \multirow{2}{*}{2} & \multirow{2}{*}{2.2075} & 0.27 \\
	&  &  & 8.38 & 395.68 &  &  & 0.15 \\
	\cline{2-8} \xrowht{5pt}
	& \multirow{2}{*}{2019-12-02 00:21:19} & \multirow{2}{*}{5159.925} & 2.25 & 110.40 & \multirow{2}{*}{2} & \multirow{2}{*}{2.2075} & 0.24 \\
	&  &  & 8.38 & 395.68 &  &  & 0.13 \\
	\cline{2-8} \xrowht{5pt}
	& \multirow{2}{*}{2019-12-02 21:20:55} & \multirow{2}{*}{5339.983} & 2.25 & 110.40 & \multirow{2}{*}{2} & \multirow{2}{*}{2.2075} & 0.25 \\
	&  &  & 8.38 & 395.68 &  &  & 0.13 \\
	\cline{2-8} \xrowht{5pt}
	& \multirow{2}{*}{2019-12-06 21:59:03} & \multirow{2}{*}{5339.983} & 2.25 & 110.40 & \multirow{2}{*}{2} & \multirow{2}{*}{2.2075} & 0.24 \\
	&  &  & 8.38 & 395.68 &  &  & 0.13 \\
	\cline{2-8} \xrowht{5pt}
	& \multirow{2}{*}{2019-12-10 20:47:43} & \multirow{2}{*}{5339.983} & 2.25 & 110.40 & \multirow{2}{*}{2} & \multirow{2}{*}{2.2075} & 0.25 \\
	&  &  & 8.38 & 395.68 &  &  & 0.13 \\
	\cline{2-8} \xrowht{5pt}
	& \multirow{2}{*}{2019-12-13 01:03:31} & \multirow{2}{*}{5119.868} & 2.25 & 110.40 & \multirow{2}{*}{2} & \multirow{2}{*}{2.2075} & 0.24 \\
	&  &  & 8.38 & 395.68 &  &  & 0.13 \\
	\cline{2-8} \xrowht{5pt}
	& \multirow{2}{*}{2019-12-16 22:37:57} & \multirow{2}{*}{10499.935} & 2.26 & 105.30 & \multirow{2}{*}{2} & \multirow{2}{*}{2.2075} & 0.25 \\
	&  &  & 8.38 & 395.68 &  &  & 0.13 \\
	\cline{2-8} \xrowht{5pt}
	& \multirow{2}{*}{2019-12-21 00:28:03} & \multirow{2}{*}{6119.969} & 2.26 & 105.30 & \multirow{2}{*}{2} & \multirow{2}{*}{2.2075} & 0.25 \\
	&  &  & 8.38 & 395.68 &  &  & 0.13 \\
	\cline{2-8} \xrowht{5pt}
	& \multirow{2}{*}{2020-02-10 00:35:01} & \multirow{2}{*}{4199.929} & 2.25 & 110.40 & \multirow{2}{*}{2} & \multirow{2}{*}{2.2075} & 0.25 \\
	&  &  & 8.38 & 395.68 &  &  & 0.14
	\enddata
	\tablecomments{FRB~121102 was observed for a total of 27.3\,hr between 2019~September~19 and 2020~February~11. During each of the two epochs highlighted in bold, 1~radio burst was detected in the 2.3\,GHz frequency band using DSS-63, but there was no evidence of radio emission at 8.4\,GHz. \\
		$^{\mathrm{a}}$ Deep Space Network~(DSN) radio antenna used for observations. \\
		$^{\mathrm{b}}$ Start time of the radio observations in yyyy-mm-dd hh:mm:ss format. \\
		$^{\mathrm{c}}$ Total usable bandwidth after radio frequency interference~(RFI) mitigation. \\
		$^{\mathrm{d}}$ Number of circular polarizations recorded. \\
		$^{\mathrm{e}}$ 6$\sigma$~fluence detection threshold, $\mathcal{F_\text{min}}$, for an assumed burst width of 1\,ms.
	}
	\label{Table:Table1}
	
\end{deluxetable*}


\newpage

\setcounter{table}{1}

\begin{deluxetable*}{c|ccccccc}
	
	\tablenum{2}
	\tabletypesize{\small}
	\tablecolumns{8}
	\tablewidth{0pt}
	\tablecaption{\textsc{Multiwavelength Radio Observations of FRB~180916.J0158+65}}
	\tablehead{
		Telescope$^{\mathrm{a}}$ & 
		Start Time$^{\mathrm{b}}$ & 
		Exposure &
		Center & 
		Bandwidth$^{\mathrm{c}}$ &
		Number of &
		Time &
		6$\sigma$ Fluence \\
		& 
		&
		Time & 
		Frequency & 
		&
		Polarizations$^{\mathrm{d}}$ &
		Resolution & 
		Detection Threshold ($\mathcal{F_\text{min}}$)$^{\mathrm{e}}$ \\
		&
		(UTC) &
		(s) &
		(GHz) & 
		(MHz) & 
		& 
		(ms) & 
		$\big(\text{Jy\,ms}\,\sqrt{(w/1\,\text{ms})}\big)$
	}
	\startdata
	\multirow{20}{*}{DSS-63} & \multirow{2}{*}{2019-09-19 02:19:37} & \multirow{2}{*}{12899.948} & 2.25 & 120.14 & 1 & \multirow{2}{*}{2.2075} & 0.33 \\
	&  &  & 8.38 & 395.68 & 2 &  & 0.13 \\
	\cline{2-8} \xrowht{5pt}
	& \multirow{2}{*}{2019-09-24 01:41:47} & \multirow{2}{*}{14579.937} & 2.25 & 120.14 & 1 & \multirow{2}{*}{2.2075} & 0.33 \\
	&  &  & 8.38 & 395.68 & 2 &  & 0.13 \\
	\cline{2-8} \xrowht{5pt}
	& \multirow{2}{*}{2019-09-28 01:37:03} & \multirow{2}{*}{11279.991} & 2.25 & 120.14 & 1 & \multirow{2}{*}{2.2075} & 0.33 \\
	&  &  & 8.38 & 395.68 & 2 &  & 0.13 \\
	\cline{2-8} \xrowht{5pt}
	& \multirow{2}{*}{2019-09-29 01:32:01} & \multirow{2}{*}{11219.933} & 2.25 & 120.14 & 1 & \multirow{2}{*}{2.2075} & 0.33 \\
	&  &  & 8.38 & 395.68 & 2 &  & 0.13 \\
	\cline{2-8} \xrowht{5pt}
	& \multirow{2}{*}{2019-10-19 22:24:39} & \multirow{2}{*}{4058.385} & 2.25 & 120.14 & 1 & \multirow{2}{*}{0.2759} & 0.33 \\
	&  &  & 8.38 & 395.68 & 2 &  & 0.13 \\
	\cline{2-8} \xrowht{5pt}
	& \multirow{2}{*}{2019-11-20 20:12:03} & \multirow{2}{*}{5639.994} & 2.25 & 110.40 & \multirow{2}{*}{2} & \multirow{2}{*}{2.2075} & 0.25 \\
	&  &  & 8.38 & 395.68 &  &  & 0.13 \\
	\cline{2-8} \xrowht{5pt}
	& \multirow{2}{*}{2019-11-21 01:29:07} & \multirow{2}{*}{5639.992} & 2.25 & 110.40 & \multirow{2}{*}{2} & \multirow{2}{*}{2.2075} & 0.25 \\
	&  &  & 8.38 & 395.68 &  &  & 0.13 \\
	\cline{2-8} \xrowht{5pt}
	& \multirow{2}{*}{2019-11-25 23:27:23} & \multirow{2}{*}{11279.991} & 2.25 & 110.40 & \multirow{2}{*}{2} & \multirow{2}{*}{2.2075} & 0.24 \\
	&  &  & 8.38 & 395.68 &  &  & 0.13 \\
	\cline{2-8} \xrowht{5pt}
	& \multirow{2}{*}{2019-12-02 02:10:35$^{\mathrm{f}}$} & \multirow{2}{*}{5939.934} & 2.25 & 110.40 & \multirow{2}{*}{2} & \multirow{2}{*}{2.2075} & 0.25 \\
	&  &  & 8.38 & 395.68 &  &  & 0.14 \\
	\cline{1-8} \xrowht{5pt}
	\multirow{2}{*}{DSS-14} & 2019-12-02 05:37:29$^{\mathrm{f}}$ & 5039.997 & 1.54 & 250.00 & 1 & 0.1024 & 0.29 \\
	\cline{2-8} \xrowht{5pt}
	& 2019-12-02 07:09:03$^{\mathrm{f}}$ & 2474.353 & 1.54 & 250.00 & 1 & 0.1024 & 0.29 \\
	\cline{1-8} \xrowht{5pt}
	\multirow{16}{*}{DSS-63} & \multirow{2}{*}{2019-12-02 22:58:41$^{\mathrm{f}}$} & \multirow{2}{*}{5939.934} & 2.25 & 110.40 & \multirow{2}{*}{2} & \multirow{2}{*}{2.2075} & 0.24 \\
	&  &  & 8.38 & 395.68 &  &  & 0.13 \\
	\cline{2-8} \xrowht{5pt}
	& \multirow{2}{*}{2019-12-03 02:32:47$^{\mathrm{f}}$} & \multirow{2}{*}{4242.772} & 2.25 & 110.40 & \multirow{2}{*}{2} & \multirow{2}{*}{2.2075} & 0.24 \\
	&  &  & 8.38 & 395.68 &  &  & 0.13 \\
	\cline{2-8} \xrowht{5pt}
	& \multirow{2}{*}{2019-12-07 00:20:21} & \multirow{2}{*}{9519.922} & 2.25 & 110.40 & \multirow{2}{*}{2} & \multirow{2}{*}{2.2075} & 0.25 \\
	&  &  & 8.38 & 395.68 &  &  & 0.13 \\
	\cline{2-8} \xrowht{5pt}
	& \multirow{2}{*}{2019-12-10 22:30:27} & \multirow{2}{*}{5639.992} & 2.25 & 110.40 & \multirow{2}{*}{2} & \multirow{2}{*}{2.2075} & 0.24 \\
	&  &  & 8.38 & 395.68 &  &  & 0.13 \\
	\cline{2-8} \xrowht{5pt}
	& \multirow{2}{*}{2019-12-12 23:21:31} & \multirow{2}{*}{5339.983} & 2.25 & 110.40 & \multirow{2}{*}{2} & \multirow{2}{*}{2.2075} & 0.26 \\
	&  &  & 8.38 & 395.68 &  &  & 0.14 \\
	\cline{2-8} \xrowht{5pt}
	& \multirow{2}{*}{2019-12-13 02:47:41} & \multirow{2}{*}{4079.952} & 2.25 & 110.40 & \multirow{2}{*}{2} & \multirow{2}{*}{2.2075} & 0.26 \\
	&  &  & 8.38 & 395.68 &  &  & 0.14 \\
	\cline{2-8} \xrowht{5pt}
	& \multirow{2}{*}{2019-12-16 20:53:47} & \multirow{2}{*}{5339.983} & 2.26 & 105.30 & \multirow{2}{*}{2} & \multirow{2}{*}{2.2075} & 0.25 \\
	&  &  & 8.38 & 395.68 &  &  & 0.13 \\
	\cline{1-8} \xrowht{5pt}
	\multirow{7}{*}{DSS-14} & 2019-12-18 02:22:05$^{\mathrm{f}}$ & 5339.994 & 1.54 & 250.00 & 1 & 0.1024 & 0.29 \\
	\cline{2-8} \xrowht{5pt}
	& 2019-12-18 04:01:13$^{\mathrm{f}}$ & 5639.994 & 1.54 & 250.00 & 1 & 0.1024 & 0.29 \\
	\cline{2-8} \xrowht{5pt}
	& 2019-12-18 06:05:43$^{\mathrm{f}}$ & 4619.994 & 1.54 & 250.00 & 1 & 0.1024 & 0.29 \\
	\cline{2-8} \xrowht{5pt}
	& 2019-12-18 07:30:17$^{\mathrm{f}}$ & 5639.994 & 1.54 & 250.00 & 1 & 0.1024 & 0.29 \\
	\cline{2-8} \xrowht{5pt}
	& 2019-12-18 09:11:47$^{\mathrm{f}}$ & 5639.994 & 1.54 & 250.00 & 1 & 0.1024 & 0.29 \\
	\cline{2-8} \xrowht{5pt}
	& 2019-12-18 10:53:33$^{\mathrm{f}}$ & 5099.992 & 1.54 & 250.00 & 1 & 0.1024 & 0.29 \\
	\cline{1-8} \xrowht{5pt}
	\multirow{34}{*}{DSS-63} & \multirow{2}{*}{2019-12-18 20:53:49$^{\mathrm{f}}$} & \multirow{2}{*}{5639.992} & 2.26 & 105.30 & \multirow{2}{*}{2} & \multirow{2}{*}{2.2075} & 0.26 \\
	&  &  & 8.38 & 395.68 &  &  & 0.14 \\
	\cline{2-8} \xrowht{5pt}
	& \multirow{2}{*}{2019-12-18 22:36:31$^{\mathrm{f}}$} & \multirow{2}{*}{5639.994} & 2.26 & 105.30 & \multirow{2}{*}{2} & \multirow{2}{*}{2.2075} & 0.26 \\
	&  &  & 8.38 & 395.68 &  &  & 0.14 \\
	\cline{2-8} \xrowht{5pt}
	& \multirow{2}{*}{2019-12-19 00:18:27$^{\mathrm{f}}$} & \multirow{2}{*}{5639.992} & 2.26 & 105.30 & \multirow{2}{*}{2} & \multirow{2}{*}{2.2075} & 0.26 \\
	&  &  & 8.38 & 395.68 &  &  & 0.14 \\
	\cline{2-8} \xrowht{5pt}
	& \multirow{2}{*}{2019-12-19 02:00:35$^{\mathrm{f}}$} & \multirow{2}{*}{6159.892} & 2.26 & 105.30 & \multirow{2}{*}{2} & \multirow{2}{*}{2.2075} & 0.26 \\
	&  &  & 8.38 & 395.68 &  &  & 0.14 \\
	\cline{2-8} \xrowht{5pt}
	& \multirow{2}{*}{2019-12-20 20:58:49} & \multirow{2}{*}{5639.994} & 2.26 & 105.30 & \multirow{2}{*}{2} & \multirow{2}{*}{2.2075} & 0.25 \\
	&  &  & 8.38 & 395.68 &  &  & 0.13 \\
	\cline{2-8} \xrowht{5pt}
	& \multirow{2}{*}{2019-12-20 22:40:45} & \multirow{2}{*}{5639.992} & 2.26 & 105.30 & \multirow{2}{*}{2} & \multirow{2}{*}{2.2075} & 0.25 \\
	&  &  & 8.38 & 395.68 &  &  & 0.13 \\
	\cline{2-8} \xrowht{5pt}
	& \multirow{2}{*}{2020-01-31 23:01:31} & \multirow{2}{*}{5379.892} & 2.25 & 110.40 & \multirow{2}{*}{2} & \multirow{2}{*}{2.2075} & 0.26 \\
	&  &  & 8.38 & 395.68 &  &  & 0.14 \\
	\cline{2-8} \xrowht{5pt}
	& \multirow{2}{*}{2020-02-02 00:16:33} & \multirow{2}{*}{11579.930} & 2.25 & 110.40 & \multirow{2}{*}{2} & \multirow{2}{*}{2.2075} & 0.27 \\
	&  &  & 8.38 & 395.68 &  &  & 0.15 \\
	\cline{2-8} \xrowht{5pt}
	& \multirow{2}{*}{2020-02-03 00:36:33} & \multirow{2}{*}{4679.973} & 2.25 & 110.40 & \multirow{2}{*}{2} & \multirow{2}{*}{2.2075} & 0.27 \\
	&  &  & 8.38 & 395.68 &  &  & 0.15 \\
	\cline{2-8} \xrowht{5pt}
	& \multirow{2}{*}{2020-02-04 00:26:33} & \multirow{2}{*}{5039.973} & 2.25 & 110.40 & \multirow{2}{*}{2} & \multirow{2}{*}{2.2075} & 0.27 \\
	&  &  & 8.38 & 395.68 &  &  & 0.15 \\
	\cline{2-8} \xrowht{5pt}
	& \multirow{2}{*}{2020-02-05 20:11:19} & \multirow{2}{*}{7499.864} & 2.25 & 110.40 & \multirow{2}{*}{2} & \multirow{2}{*}{2.2075} & 0.25 \\
	&  &  & 8.38 & 395.68 &  &  & 0.13 \\
	\cline{2-8} \xrowht{5pt}
	& \multirow{2}{*}{2020-02-06 22:01:27} & \multirow{2}{*}{12779.971} & 2.25 & 110.40 & \multirow{2}{*}{2} & \multirow{2}{*}{2.2075} & 0.26 \\
	&  &  & 8.38 & 395.68 &  &  & 0.14 \\
	\cline{2-8} \xrowht{5pt}
	& \multirow{2}{*}{2020-02-07 23:01:33} & \multirow{2}{*}{8819.990} & 2.25 & 110.40 & \multirow{2}{*}{2} & \multirow{2}{*}{2.2075} & 0.26 \\
	&  &  & 8.38 & 395.68 &  &  & 0.15 \\
	\cline{2-8} \xrowht{5pt}
	& \multirow{2}{*}{2020-02-09 21:51:29} & \multirow{2}{*}{9239.977} & 2.25 & 110.40 & \multirow{2}{*}{2} & \multirow{2}{*}{2.2075} & 0.26 \\
	&  &  & 8.38 & 395.68 &  &  & 0.14 \\
	\cline{2-8} \xrowht{5pt}
	& \multirow{2}{*}{2020-02-10 22:06:29} & \multirow{2}{*}{11399.946} & 2.25 & 110.40 & \multirow{2}{*}{2} & \multirow{2}{*}{2.2075} & 0.26 \\
	&  &  & 8.38 & 395.68 &  &  & 0.14
	\enddata
	\vspace{-0.09cm}
	\label{Table:Table2}
	
\end{deluxetable*}

\clearpage

\setcounter{page}{14}
\discardpages{14}


\newpage

\begin{deluxetable*}{c|ccccccc}
	
	\tablenum{2}
	\tabletypesize{\small}
	\tablecolumns{8}
	\tablewidth{0pt}
	\tablecaption{\vspace{-0.3cm}\hspace{2.8cm}--- \textit{Continued} \\ \textsc{Multiwavelength Radio Observations of FRB~180916.J0158+65 --- \textit{Continued}}}
	\tablehead{
		Telescope$^{\mathrm{a}}$ & 
		Start Time$^{\mathrm{b}}$ & 
		Exposure &
		Center & 
		Bandwidth$^{\mathrm{c}}$ &
		Number of &
		Time & 
		6$\sigma$ Fluence \\
		& 
		&
		Time & 
		Frequency & 
		&
		Polarizations$^{\mathrm{d}}$ &
		Resolution & 
		Detection Threshold ($\mathcal{F_\text{min}}$)$^{\mathrm{e}}$ \\
		&
		(UTC) &
		(s) &
		(GHz) & 
		(MHz) & 
		& 
		(ms) & 
		$\big(\text{Jy\,ms}\,\sqrt{(w/1\,\text{ms})}\big)$
	}
	\startdata
	\multirow{20}{*}{DSS-63} & \multirow{2}{*}{2020-03-26 10:58:31} & \multirow{2}{*}{8919.892} & 2.25 & 110.40 & \multirow{2}{*}{2} & \multirow{2}{*}{2.2075} & 0.24 \\
	&  &  & 8.38 & 395.68 &  &  & 0.13 \\
	\cline{2-8} \xrowht{5pt}
	& \multirow{2}{*}{2020-04-03 11:30:01} & \multirow{2}{*}{3299.942} & 2.25 & 110.40 & \multirow{2}{*}{2} & \multirow{2}{*}{2.2075} & 0.25 \\
	&  &  & 8.38 & 395.68 &  &  & 0.13 \\
	\cline{2-8} \xrowht{5pt}
	& \multirow{2}{*}{2020-04-03 15:20:01} & \multirow{2}{*}{9359.929} & 2.25 & 110.40 & \multirow{2}{*}{2} & \multirow{2}{*}{2.2075} & 0.25 \\
	&  &  & 8.38 & 395.68 &  &  & 0.13 \\
	\cline{2-8} \xrowht{5pt}
	& \multirow{2}{*}{2020-04-05 18:26:29} & \multirow{2}{*}{11579.928} & 2.25 & 110.40 & \multirow{2}{*}{2} & \multirow{2}{*}{2.2075} & 0.26 \\
	&  &  & 8.38 & 395.68 &  &  & 0.14 \\
	\cline{2-8} \xrowht{5pt}
	& \multirow{2}{*}{2020-04-08 09:53:11} & \multirow{2}{*}{14579.953} & 2.25 & 110.40 & \multirow{2}{*}{2} & \multirow{2}{*}{2.2075} & 0.24 \\
	&  &  & 8.38 & 395.68 &  &  & 0.13 \\
	\cline{2-8} \xrowht{5pt}
	& \multirow{2}{*}{2020-04-15 11:12:29} & \multirow{2}{*}{9778.488} & 2.25 & 110.40 & \multirow{2}{*}{2} & \multirow{2}{*}{1.0240} & 0.24 \\
	&  &  & 8.38 & 395.68 &  &  & 0.13 \\
	\cline{2-8} \xrowht{5pt}
	& \multirow{2}{*}{2020-05-03 12:18:47} & \multirow{2}{*}{26278.674} & 2.25 & 110.40 & \multirow{2}{*}{2} & \multirow{2}{*}{1.0240} & 0.24 \\
	&  &  & 8.38 & 395.68 &  &  & 0.13 \\
	\cline{2-8} \xrowht{5pt}
	& \multirow{2}{*}{2020-05-10 17:45:01} & \multirow{2}{*}{5924.631} & 2.25 & 110.40 & \multirow{2}{*}{2} & \multirow{2}{*}{1.0240} & 0.27 \\
	&  &  & 8.38 & 395.68 &  &  & 0.15 \\
	\cline{2-8} \xrowht{5pt}
	& \multirow{2}{*}{2020-05-13 18:05:01} & \multirow{2}{*}{3719.920} & 2.25 & 110.40 & \multirow{2}{*}{2} & \multirow{2}{*}{1.0240} & 0.27 \\
	&  &  & 8.38 & 395.68 &  &  & 0.15
	\enddata
	\vspace{-0.09cm}
	\tablecomments{FRB~180916.J0158+65 was observed for a total of 101.8\,hr between 2019~September~19 and 2020~May~14. No radio bursts were detected with DSS-63/DSS-14 in any frequency band during these epochs. \\
		$^{\mathrm{a}}$ Deep Space Network~(DSN) radio antenna used for observations. \\
		$^{\mathrm{b}}$ Start time of the radio observations in yyyy-mm-dd hh:mm:ss format. \\
		$^{\mathrm{c}}$ Total usable bandwidth after radio frequency interference~(RFI) mitigation. \\
		$^{\mathrm{d}}$ Number of circular polarizations recorded. \\
		$^{\mathrm{e}}$ 6$\sigma$~fluence detection threshold, $\mathcal{F_\text{min}}$, for an assumed burst width of 1\,ms. \\
		$^{\mathrm{f}}$ These radio observations were also presented in~\citet{Scholz+2020}.
	}
	\label{Table:Table2Continued}
	
\end{deluxetable*}

\clearpage

\setcounter{page}{15}
\discardpages{15}


\newpage

\setcounter{table}{2}

\setlength{\evensidemargin}{-0.6cm}

\begin{deluxetable*}{ccccccccc}
	
	\tablenum{3}
	\tabletypesize{\small}
	\tablecolumns{9}
	\tablewidth{0pt}
	\tablecaption{\textsc{Radio Bursts Detected from FRB~121102 with DSS-63}}
	\tablehead{
		\colhead{Burst ID} & 
		\colhead{Peak Time$^{\mathrm{a,i}}$} & 
		\colhead{(S/N)$_{\text{peak}}$$^{\mathrm{b,i}}$} & 
		\colhead{DM$^{\mathrm{c}}$} &
		\colhead{Burst Width$^{\mathrm{d,i}}$} &
		\colhead{Peak Flux Density$^{\mathrm{e,i}}$} & 
		\colhead{Fluence ($\mathcal{F}$)$^{\mathrm{f,i}}$} &
		\colhead{Spectral Energy Density$^{\mathrm{g,i}}$} &
		\colhead{Isotropic-Equivalent Energy$^{\mathrm{h,i}}$} \\
		\colhead{} &
		\colhead{(MJD)} &
		\colhead{} & 
		\colhead{(pc\,cm$^{\text{--3}}$)} & 
		\colhead{(ms)} & 
		\colhead{(Jy)} &
		\colhead{(Jy\,ms)} &
		\colhead{(10$^{\text{30}}$\,erg\,Hz$^{\text{--1}}$)} &
		\colhead{(10$^{\text{38}}$\,erg)}
	}
	\startdata
	B1 & 58754.04745852 & 40.26 & 563.0 & 2.4\,$\pm$\,0.1 & 1.6\,$\pm$\,0.3 & 3.3\,$\pm$\,0.5 & 3.2\,$\pm$\,0.5 & 3.8\,$\pm$\,0.6 \\
	B2 & 58755.03200822 & 37.27 & 563.0 & 2.6\,$\pm$\,0.2 & 1.5\,$\pm$\,0.3 & 4.6\,$\pm$\,0.7 & 3.8\,$\pm$\,0.5 & 4.5\,$\pm$\,0.6
	\enddata
	\tablecomments{These $S$-band radio bursts were detected in data recorded with a time resolution of 2.2\,ms. We were not sensitive to bursts with narrower widths than the sampling time of our observations~(see Table~\ref{Table:Table1}). \\
		$^{\mathrm{a}}$ Barycentric time at the peak of the burst, determined after removing the time delay from dispersion using a DM of 563.0\,pc\,cm$^{\text{--3}}$ and correcting to infinite frequency. The barycentric times were derived using the position ($\alpha_{\text{J2000}}$\,$=$\,05$^{\text{h}}$31$^{\text{m}}$58$^{\text{s}}$.698, $\delta_{\text{J2000}}$\,$=$\,33$^{\circ}$08$\arcmin$52$\arcsec$.586) in \citet{Marcote+2017}. \\
		$^{\mathrm{b}}$ Peak signal-to-noise ratio,~(S/N)$_{\text{peak}}$. \\
		$^{\mathrm{c}}$ Average DM near the time of each burst (A. D. Seymour, private communication). \\
		$^{\mathrm{d}}$ Full width at half maximum (FWHM) temporal duration, determined from a Gaussian fit to the dedispersed burst profile. \\
		$^{\mathrm{e}}$ Uncertainties are dominated by the 20\% fractional error on the system temperature, $T_{\text{sys}}$. \\
		$^{\mathrm{f}}$ Time-integrated burst fluence ($\mathcal{F}$), determined using the 2$\sigma$ FWHM for the duration of the burst. \\
		$^{\mathrm{g}}$ Spectral energy density values were calculated assuming isotropic emission and using the expression $4\pi d_{L}^2 \mathcal{F} /(1$\,$+$\,$z)$, where $d_{L}$\,$=$\,972\,Mpc is the luminosity distance of FRB~121102~\citep{Tendulkar+2017}, $\mathcal{F}$ is the burst fluence, and $z$\,$=$\,0.19273(8) is the redshift of the dwarf host galaxy~\citep{Tendulkar+2017}. \\
		$^{\mathrm{h}}$ Isotropic-equivalent energy values were calculated for a bandwidth of 118.75\,MHz, which corresponds to the usable portion of the 2.25\,GHz frequency band after RFI mitation at the time of the burst. \\
		$^{\mathrm{i}}$ Values were derived after dedispersing each burst using a DM of 563.0\,pc\,cm$^{\text{--3}}$.}
	\label{Table:Table3}
	
\end{deluxetable*}

\clearpage


\newpage

\setcounter{table}{3}

\begin{deluxetable*}{cccccc}
	
	\tablenum{4}
	\tabletypesize{\small}
	\tablecolumns{6}
	\tablewidth{0pt}
	\tablecaption{\textsc{Radio Bursts Detected from FRB~180916.J0158+65 with CHIME/FRB \\ During Simultaneous Radio Observations with DSS-63/DSS-14}}
	\tablehead{
		\colhead{Burst ID} & 
		\colhead{Peak Time$^{\mathrm{a}}$} &
		\colhead{DM} &
		\colhead{Burst Width} &
		\colhead{Peak Flux Density} & 
		\colhead{Fluence ($\mathcal{F}$)} \\
		\colhead{} &
		\colhead{(MJD)} &
		\colhead{(pc\,cm$^{\text{--3}}$)} & 
		\colhead{(ms)} & 
		\colhead{(Jy)} &
		\colhead{(Jy\,ms)}
	}
	\startdata
	C1 & 58835.17721035 & 349.5\,$\pm$\,0.5 & 5.0\,$\pm$\,0.5 & 0.4\,$\pm$\,0.2 & 2.9\,$\pm$\,0.7 \\
	C2 & 58882.04838586 & 349.4\,$\pm$\,0.3 & 1.14\,$\pm$\,0.12 & $>$\,0.5\,$\pm$\,0.2 & $>$\,0.8\,$\pm$\,0.3 \\
	C3 & 58883.02020163 & 370.4\,$\pm$\,1.6 \\
	C4 & 58883.04146680 & 349.6\,$\pm$\,0.3 & 8.6\,$\pm$\,0.5 & $>$\,0.4\,$\pm$\,0.3 & $>$\,4.3\,$\pm$\,1.6 \\
	C5 & 58883.04307123 & 349.81\,$\pm$\,0.05 & 1.157\,$\pm$\,0.011 & 6.1\,$\pm$\,2.0& 16.3\,$\pm$\,5.0 \\
	C6 & 58883.04556977 & 349.8\,$\pm$\,0.5 & 1.48\,$\pm$\,0.13 & 0.5\,$\pm$\,0.2 & 1.5\,$\pm$\,0.6 \\
	C7 & 58883.05523556 & 348.7\,$\pm$\,0.6 & 0.76\,$\pm$\,0.07 & $>$\,0.5\,$\pm$\,0.3 & $>$\,0.4\,$\pm$\,0.1 \\
	C8 & 58982.76846813 & 352.6\,$\pm$\,3.2
	\enddata
	\tablecomments{The burst properties are reproduced from~\citet{CHIME+2020b}\textsuperscript{\ref{FootnoteCHIMEWebsite}}. The burst widths, peak flux densities, and fluences of C3 and C8 have not yet been reported. \\
		$^{\mathrm{a}}$ Barycentric time at the peak of the burst after removing the time delay from dispersion using the listed DM and correcting to infinite frequency. The barycentric times were determined using the position ($\alpha_{\text{J2000}}$\,$=$\,01$^{\text{h}}$58$^{\text{m}}$00$^{\text{s}}$.75017, $\delta_{\text{J2000}}$\,$=$\,65$^{\circ}$43$\arcmin$00$\arcsec$.3152) in \citet{Marcote+2020}.}
	\label{Table:Table4}
	
\end{deluxetable*}


\setcounter{table}{4}

\begin{deluxetable*}{ccccccc}
	
	\tablenum{5}
	\tabletypesize{\small}
	\tablecolumns{7}
	\tablewidth{0pt}
	\tablecaption{\textsc{Estimated Number of Radio Bursts Detectable from FRB~180916.J0158+65}}
	\tablehead{
		\colhead{Time Interval} & 
		\colhead{Burst Rate ($r$)} &
		\colhead{Number of Bursts ($N$)} &
		\colhead{Burst Rate ($r$)} &
		\colhead{Number of Bursts ($N$)} & 
		\colhead{Burst Rate ($r$)} &
		\colhead{Number of Bursts ($N$)} \\
		\colhead{} &
		\colhead{at 1.5\,GHz$^{\mathrm{a}}$} &
		\colhead{Detectable at 1.5\,GHz$^{\mathrm{b}}$} &
		\colhead{at 2.3\,GHz$^{\mathrm{a}}$} &
		\colhead{Detectable at 2.3\,GHz$^{\mathrm{b}}$} & 
		\colhead{at 8.4\,GHz$^{\mathrm{a}}$} &
		\colhead{Detectable at 8.4\,GHz$^{\mathrm{b}}$} \\
		\colhead{} &
		\colhead{(Bursts Per Hour)} &
		\colhead{(Bursts)} & 
		\colhead{(Bursts Per Hour)} &
		\colhead{(Bursts)} &
		\colhead{(Bursts Per Hour)} &
		\colhead{(Bursts)} \\
	}
	\startdata
	$\pm$2.7\,d Activity Window$^{\mathrm{c}}$ & 1.2$^{\text{+2.8}}_{\text{--1.8}}$ &  & 0.3$^{\text{+1.0}}_{\text{--0.6}}$ & 12.1$^{\text{+37.6}}_{\text{--23.1}}$ & 0.006$^{\text{+0.038}}_{\text{--0.023}}$ & 0.2$^{\text{+1.4}}_{\text{--0.8}}$ \\
	Subinterval 1$^{\mathrm{d}}$ & 2.4$^{\text{+5.7}}_{\text{--3.6}}$ & 26.5$^{\text{+62.2}}_{\text{--39.2}}$ & 0.7$^{\text{+2.1}}_{\text{--1.3}}$ & 8.0$^{\text{+25.1}}_{\text{--15.2}}$ & 0.01$^{\text{+0.08}}_{\text{--0.05}}$ & 0.1$^{\text{+0.9}}_{\text{--0.5}}$ \\
	Subinterval 2$^{\mathrm{e}}$ & 1.1$^{\text{+2.7}}_{\text{--1.7}}$ &  & 0.3$^{\text{+1.0}}_{\text{--0.6}}$ & 4.7$^{\text{+15.4}}_{\text{--9.2}}$ & 0.006$^{\text{+0.034}}_{\text{--0.020}}$ & 0.09$^{\text{+0.54}}_{\text{--0.32}}$ \\
	Subinterval 3$^{\mathrm{f}}$ & 0.1$^{\text{+0.9}}_{\text{--0.2}}$ &  & 0.04$^{\text{+0.25}}_{\text{--0.08}}$ & 0.3$^{\text{+2.1}}_{\text{--0.7}}$ & 0.0007$^{\text{+0.0059}}_{\text{--0.0026}}$ & 0.006$^{\text{+0.050}}_{\text{--0.022}}$
	\enddata
	\tablecomments{Since the $L$-band (1.5\,GHz) observations of FRB~180916.J0158+65 with DSS-14 occurred entirely during subinterval~1, the number of detectable bursts at this frequency is only estimated in this time interval. \\
		$^{\mathrm{a}}$ Burst rates at each frequency were calculated by substituting the fluence detection thresholds reported by CHIME/FRB in each time interval~\citep{CHIME+2020b} and the average 6$\sigma$~fluence detection thresholds during our $L$-band, $S$-band, and $X$-band observations with DSS-14 and DSS-63 into Equation~\eqref{Equation:DetectionRate}, using power law indices of $\alpha_{\text{s}}$\,$=$\,\text{--1.6}$^{\text{+1.0}}_{\text{--0.6}}$~\citep{Chawla+2020} and $\gamma$\,$=$\,--2.3\,$\pm$\,0.4~\citep{CHIME+2020b}. The average 6$\sigma$~fluence detection thresholds of DSS-14 and DSS-63 at $L$-band, $S$-band, and $X$-band were 0.29, 0.26, and 0.14\,Jy\,ms, respectively. \\
		$^{\mathrm{b}}$ Expected number of detectable bursts at each frequency band above the average 6$\sigma$~fluence detection thresholds of DSS-14 and DSS-63. The number of detectable bursts was determined using the formula $N$\,$=$\,$r$\,$\times$\,$t$, where $r$ is the calculated rate of detectable bursts and $t$ is the total exposure time in each time interval. \\
		$^{\mathrm{c}}$ The $\pm$2.7\,d interval around the peak of FRB~180916.J0158+65's activity window, defined in~\citet{CHIME+2020b}, which corresponds to activity phases between $\sim$0.33--0.67. \\
		$^{\mathrm{d}}$ Subinterval~1 is defined as $\pm$0.9\,d around the activity peak, which corresponds to activity phases between \mbox{$\sim$0.44--0.56}~\citep{CHIME+2020b}. \\
		$^{\mathrm{e}}$ Subinterval~2 is defined as times between 0.9 and 1.8\,d from the activity peak, which corresponds to activity phases \mbox{$\sim$0.39--0.44} and \mbox{$\sim$0.56--0.61}~\citep{CHIME+2020b}. \\
		$^{\mathrm{f}}$ Subinterval~3 is defined as times between 1.8 and 2.7\,d from the activity peak, which corresponds to activity phases \mbox{$\sim$0.33--0.39} and \mbox{$\sim$0.61--0.67}~\citep{CHIME+2020b}.}
	\label{Table:Table5}
	
\end{deluxetable*}


\end{document}